\documentclass[a4paper,11pt]{article}
\pdfoutput=1
\usepackage{jheppub}
\usepackage{amsmath}
\usepackage{amssymb}
\usepackage{dcolumn}	
\usepackage{bm}			
\usepackage{bbm} 		
\usepackage{multirow}
\usepackage{blkarray}
\usepackage{slashed}
\usepackage{bbding}
\usepackage{pifont}
\usepackage[usenames,dvipsnames]{xcolor}
\definecolor{hgreen}{rgb}{0,.3,0}
\definecolor{hred}{rgb}{.3,0,0}
\definecolor{hblue}{rgb}{0,0,.3}
\definecolor{LightGray}{gray}{0.95}

\usepackage{adjustbox}

\graphicspath{{./Figures/}}
\makeatletter
\def\endfmffile{%
	\fmfcmd{\p@rcent\space the end.^^J%
		end.^^J%
		endinput;}%
	\if@fmfio
	\immediate\closeout\@outfmf
	\fi
	\ifnum\pdfshellescape>\z@
	\immediate\write18{mpost \thefmffile}%
	\fi}
\makeatother

\usepackage{pbox}
\usepackage{caption}
\usepackage{subcaption}
\usepackage{xcolor}
\usepackage[utf8]{inputenc}

\graphicspath{{./Figure/}}

\begin{document}


\title{Testing Lepton Flavor Universality at the Electron-Ion Collider}

\author[a]{Yongjie Deng,}
\affiliation[a]{Key Laboratory of Particle Physics and Particle Irradiation (MOE), Institute of Frontier and Interdisciplinary Science, Shandong University, Qingdao, Shandong 266237, China}
\author[b,c,1]{Xu-Hui Jiang,}
\affiliation[b]{Institute of High Energy Physics, Chinese Academy of Sciences, Beijing 100049, China}
\affiliation[c]{China Center of Advanced Science and Technology, Beijing 100190, China}
\author[a,d]{Tianbo Liu,}
\affiliation[d]{Southern Center for Nuclear-Science Theory (SCNT), Institute of Modern Physics, Chinese Academy of Sciences, Huizhou, Guangdong 516000, China}
\author[b,e,1]{Bin Yan,}
\affiliation[e]{Center for High Energy Physics, Peking University, Beijing 100871, China}
\footnotetext[1]{Corresponding author.}

\date{\today}
\emailAdd{yongjie.deng@mail.sdu.edu.cn, jiangxh@ihep.ac.cn, liutb@sdu.edu.cn, yanbin@ihep.ac.cn}

\abstract{Measurements of $b\to c\tau^-\bar\nu_\tau$ transitions at colliders are highly motivated for testing lepton flavor universality (LFU), a cornerstone hypothesis of the Standard Model (SM). Potential observations of LFU violation could provide significant evidence for physics beyond the SM (BSM). The substantial production of $b$-hadrons at the Electron-Ion Collider (EIC) would highlight its potential to support LFU testing and complement studies conducted at other experimental facilities. In this paper, we study the production of $b$-hadrons in deep inelastic scattering processes at the EIC with $\sqrt{s}=100\,\rm GeV$. We estimate the $b$-hadron yields at the EIC to reach $\mathcal O(10^9)$ with an integrated luminosity of up to $10^3\,{\rm fb}^{-1}$. Furthermore, we perform a systematic study on various $b$-hadron decays, exploring the sensitivities of LFU-violating observables, including $R_{J/\psi}$, $R_{D_s^{(\ast)}}$ and $R_{\Lambda_c}$. Detailed strategies for track-based event reconstruction are investigated for these observables. We also include a discussion of the annihilation process $B_c^+ \to \tau^+\nu_\tau$. Finally, we provide a theoretical interpretation of the projected sensitivities within the framework of low energy effective field theory (LEFT). Our analysis indicates that the EIC has the potential to constrain the relevant Wilson coefficients at $\mathcal O(0.1)$, offering complementary insights to other measurements.  
}

\maketitle

\section{Introduction\label{sec:intro}}

Lepton flavor universality (LFU) is a fundamental hypothesis of the Standard Model (SM) that requires all leptons to share identical gauge interactions. Precision testing of LFU is crucial, as any observed violation could indicate the presence of new physics beyond the SM (BSM). One of the most promising avenues for probing LFU is through the study of (semi-)leptonic decays of heavy flavor hadrons at colliders. These decays include both flavor-changing-neutral-current (FCNC) processes, such as $b\to s \ell^+\ell^-$ and $b\to s \nu\bar\nu$ transitions, as well as flavor-changing-charged-current (FCCC) processes, like $b\to c\ell\nu$. The ratios of branching fractions involving different lepton flavors serve as particularly sensitive probes of LFU, as most uncertainties associated with these rare decays cancel out in such ratios, allowing them to be calculated and measured with very high accuracy. 

For FCNC processes, the key observables are defined as
\begin{equation}~\label{eq:rkrkast}
   R_{H_s} = \frac{{\rm Br}(H_b \to H_s \mu^+ \mu^-)}{{\rm Br}(H_b \to H_s e^+ e^-)}~,
\end{equation}
where $H_b$ and $H_s$ represent hadrons containing a $b$ and $s$ quark, respectively. The most well-known instance involves the ratios $R_{K^{(\ast)}}$ , defined by setting $H_b = B$ and $H_s = K^{(\ast)}$. Recent measurements at LHCb~\cite{LHCb:2022vje} report a high consistency of these observables with SM predictions, indicating that LFU-violating effects for light leptons (electron and muon) are already highly constrained. The significant mass hierarchy between the $\tau$ lepton and the light leptons suggests that the $\tau$ lepton could be more sensitive to the BSM effects~\cite{King:2020qaj, Novichkov:2021evw}. Consequently, extending LFU studies to include the $\tau$ lepton is particularly important, such as $b\to s \tau^+ \tau^-$ transition. Unfortunately, none of these channels have been observed so far. Future lepton colliders may provide opportunities for discovery of these signals (see Ref.~\cite{Li:2020bvr} for more details), thanks to their clean collision environments and advanced $\tau-$tagging technologies. Additionally, $b \to s \nu\bar\nu$ transitions could also provide constraints on the LFU-violating effects through inclusive signal rates, although direct probing of LFU violation in these channels is challenging due to the missing energy from neutrinos in collider experiments. 

However, the FCNC processes in the SM are loop-suppressed, with the leading contributions arising from box and penguin diagrams. The FCCC processes are tree-level and thus more prominent. As a result, FCCC channels are especially important for testing LFU. A well-known example is the measurements of  $R_D$ and $R_{D^\ast}$, which have shown a significant discrepancy when compared to the SM predictions by considering the decays of $B^{(0,~\pm)}\to D^{(0,~\pm)}$ and $B^{(0,~\pm)}\to D^{\ast(0,~\pm)}$ measured before 2022~\cite{HFLAV:2022esi}. While the most recent measurements of these observables from LHCb~\cite{LHCb:2023zxo, LHCb:2023uiv, LHCb:2024jll} and Belle II~\cite{Belle-II:2023aih, Belle-II:2024ami} are consistent with the SM, potential BSM effects may still be obscured due to the significant experimental uncertainties. To further enhance the sensitivity of LFU tests, it is valuable to investigate all possible $b\to c$ transition processes and consider the ratios,
\begin{equation}
R_{H_c}\equiv \frac{{\rm Br}(H_b\to H_c\tau\nu)}{{\rm Br}(H_b\to H_c\ell\nu)}~,
\end{equation} 
where $H_b$ and $H_c$ denote hadrons containing a $b$ and $c$ quark, respectively, and $\ell=e,\mu$. The current status of these measurements is summarized in Table~\ref{tab:CStatus}. Unfortunately, these exotic decays are either still in the early stage with large uncertainties or yet to be measured. A global fit incorporating the most recent measurements up to 2024 can be found in Ref.~\cite{Iguro:2024hyk}. Additionally, the polarization from hadrons could provide further insights in probing these BSM effects. For example, the longitudinal polarization of $D^\ast$ has been measured by both the LHCb~\cite{LHCb:2023ssl} and Belle collaborations~\cite{Belle:2019ewo}, showing consistency with the SM up to 2$\sigma$ level. Although these limits could be improved by one to two orders of magnitude with an upgrade of LHCb~\cite{LHCb:2018roe} and Belle II~\cite{Belle-II:2018jsg}, they remain well above the SM predictions~\cite{ParticleDataGroup:2022pth}. Significant improvements in these measurements could be achieved in future $Z$-factories, as demonstrated in Refs.~\cite{Li:2020bvr, Ho:2022ipo, Zheng:2020ult,Zuo:2023dzn}, such as CEPC~\cite{CEPCStudyGroup:2018ghi, CEPCStudyGroup:2023quu} and FCC-$ee$~\cite{FCC:2018evy}.

\begin{table}[htbp]
\centering
\begin{tabular}{ccccc}
\hline
 & $H_b$ & $H_c$ & SM Prediction & Experimental Average\\
\hline
$R_{J/\psi}$ & $B_c$ & $J/\psi$ & $0.289$~\cite{Wang:2012lrc, Watanabe:2017mip, Asadi:2019xrc} & $0.71\pm 0.17\pm 0.18$~\cite{LHCb:2017vlu}\\
$R_{D_s}$ & $B_s$ & $D_s$ & $0.393$~\cite{Fan:2013kqa, Zhang:2022opp, Hu:2019bdf, Faustov:2012mt, Monahan:2017uby, Dutta:2018jxz, Soni:2021fky} & N/A\\
$R_{D_s^\ast}$ & $B_s$ & $D_s^\ast$ & $0.303$~\cite{Fan:2013kqa, Hu:2019bdf, Faustov:2012mt, Soni:2021fky} & N/A\\
$R_{\Lambda_c}$  & $\Lambda_b$ &  $\Lambda_c$ & $0.334$~\cite{Shivashankara:2015cta, Gutsche:2015mxa, Detmold:2015aaa, Dutta:2015ueb, Datta:2017aue} & $0.242\pm 0.076$~\cite{LHCb:2022piu}\\
\hline
\end{tabular}
\caption{The SM predictions and experimental measurements for $R_{H_c}$ observables.}\label{tab:CStatus}
\end{table}

\begin{figure}[htbp]
	\centering
	\includegraphics[width=10cm]{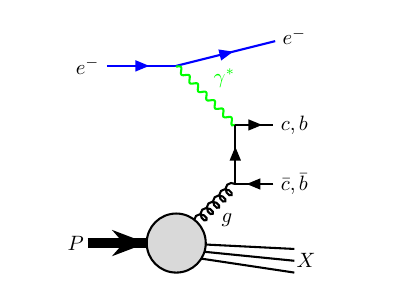}
	\caption{The Feynman diagram to illustrate the $b$-production via gluon-photon fusion in DIS processes at the EIC.} 
	\label{fig:fusion}
\end{figure}

In this paper, we extend the analysis of LFU testing at $Z$-factories and explore the potential for probing these BSM effects at the forthcoming Electron-Ion Collider (EIC). Although the EIC was initially designed to precisely determine  nuclear structures~\cite{AbdulKhalek:2021gbh}, as discussed in Ref.~\cite{Li:2023avn}, it also serves as a powerful flavor machine and shows great capabilities in probing the electroweak properties of the SM and beyond~\cite{Gonderinger:2010yn,Boughezal:2020uwq,Cirigliano:2021img,Yan:2021htf,Liu:2021lan,Li:2021uww,Davoudiasl:2021mjy,Yan:2022npz,Zhang:2022zuz,Batell:2022ogj,Boughezal:2022pmb,AbdulKhalek:2022hcn,Davoudiasl:2023pkq,Boughezal:2023ooo,Delzanno:2024ooj,Wang:2024zns,Wen:2024cfu,Du:2024sjt,Gao:2024rgl,Balkin:2023gya}.
Given that maximizing the integrated luminosity $\mathcal{L}$ has a more significant impact on the sensitivity for probing these BSM effects, compared to a slight increase in collider energy $\sqrt{s}$,  we consider beam energies of $E_e=10 ~\rm GeV$ and $E_p=250 ~\rm GeV$, corresponding to $\sqrt{s}\sim 100 ~\rm GeV$ in this study. This configuration is expected to achieve the highest integrated luminosity~\cite{AbdulKhalek:2021gbh}, and we adopt $\mathcal{L}=1000~{\rm fb}^{-1}$ as a benchmark. The production of $b$ hadrons at the EIC occurs through the gluon-photon fusion channel in deep inelastic scattering (DIS) process~\cite{Hu:2024qre}, as illustrated in Fig.~\ref{fig:fusion}. We evaluate the $b$ hadron production rates at the leading order (LO) and summarize the expected event numbers in Table~\ref{tab:brate}. It shows that the production events of $B^{(0,~\pm)}$ at the EIC is expected to reach $\sim 1.2\times 10^{9}$, which is about one order of magnitude lower than that at Belle II. However, the EIC is not constrained by the energy threshold limitations, allowing for the abundant production of heavier $b$-mesons, such as   $B_c^\pm$, as well as $b$-baryons like $\Lambda_b (\bar\Lambda_b)$. The production rates at the EIC are roughly four or five orders of magnitude lower than those at LHCb, but the EIC benefits from a cleaner collision environment in electron-proton collisions, which could significantly enhance event reconstruction and signal-background classification. More importantly, the EIC's beam polarization provides a unique advantage, enhancing sensitivity to certain BSM parameters and providing complementary constraints that are not accessible in other facilities.

\begin{table}
\centering
\begin{tabular}{cccccc}
\hline 
   & Belle~II & LHCb & Tera-$Z$ & EIC  \\ 
\hline 
$B^0$, $\bar{B}^0$ & $5.3\times 10^{10}$ & $ 6\times 10^{13}$  & $1.2 \times 10^{11}$ & $1.2 \times 10^{9}$\\
$B^\pm$ & $5.6\times 10^{10}$ & $ 6\times 10^{13}$  & $1.2 \times 10^{11}$ & $1.2 \times 10^{9}$ \\
$B_s$, $\bar{B}_s$ & $5.7 \times 10^{8}$ & $ 2\times 10^{13}$  & $3.1\times 10^{10}$ & $3.2\times 10^{8}$ \\
$B_c^\pm$ & - & $ 4 \times 10^{11}$  & $1.8\times 10^8$ & $2.4\times 10^6$ \\
$\Lambda_b$, $\bar{\Lambda}_b$ & - & $ 2\times 10^{13}$  & $2.5\times 10^{10}$ & $6.2\times 10^{8}$ \\
\hline
\end{tabular}
\caption{Expected $b$-hadron yields at Belle~II~\cite{Belle-II:2018jsg}, LHCb Upgrade II~\cite{LHCb:2018roe}, Tera-$Z$ and EIC. The yields for the first three facilities are taken from Ref.~\cite{Ai:2024nmn}, while the production rates at the EIC are estimated in Appendix~\ref{app: bhadrons}.}\label{tab:brate}
\end{table}

This paper is organized as follows. In Sec.~\ref{sec: generation}, we provide a general introduction to the strategies employed for generating signal and background events at the EIC. The analysis of $R_{J/\psi}$, $R_{D_s^{(\ast)}}$ and $R_{\Lambda_c}$, as well as the evaluation of $B_c^{\pm}$ annihilating into $\tau$ lepton and neutrino will be presented in Sec.~\ref{sec: processes}. In Sec.~\ref{sec: eft}, we present the projected limits within the framework of Low-Energy Effective Field Theory (LEFT). Finally, we summarize our findings and conclusions in Sec.~\ref{sec: conclu}.

\section{Strategy of event generation\label{sec: generation}}
 
In this study, we focus on the muon decay channel  $H_b \to H_c \mu \nu $ at the EIC to avoid potential misidentification between the electron originating from the DIS process and the leptonic decay products of  $H_b$. It is important to note that, when measuring the ratio $R_{H_c}$, the $H_b\to H_c \tau\nu$ and $H_b\to H_c \mu\nu$ decay modes act as mutual backgrounds in their respective measurements.

\begin{figure}[htbp]
	\centering
	\includegraphics[width=10cm]{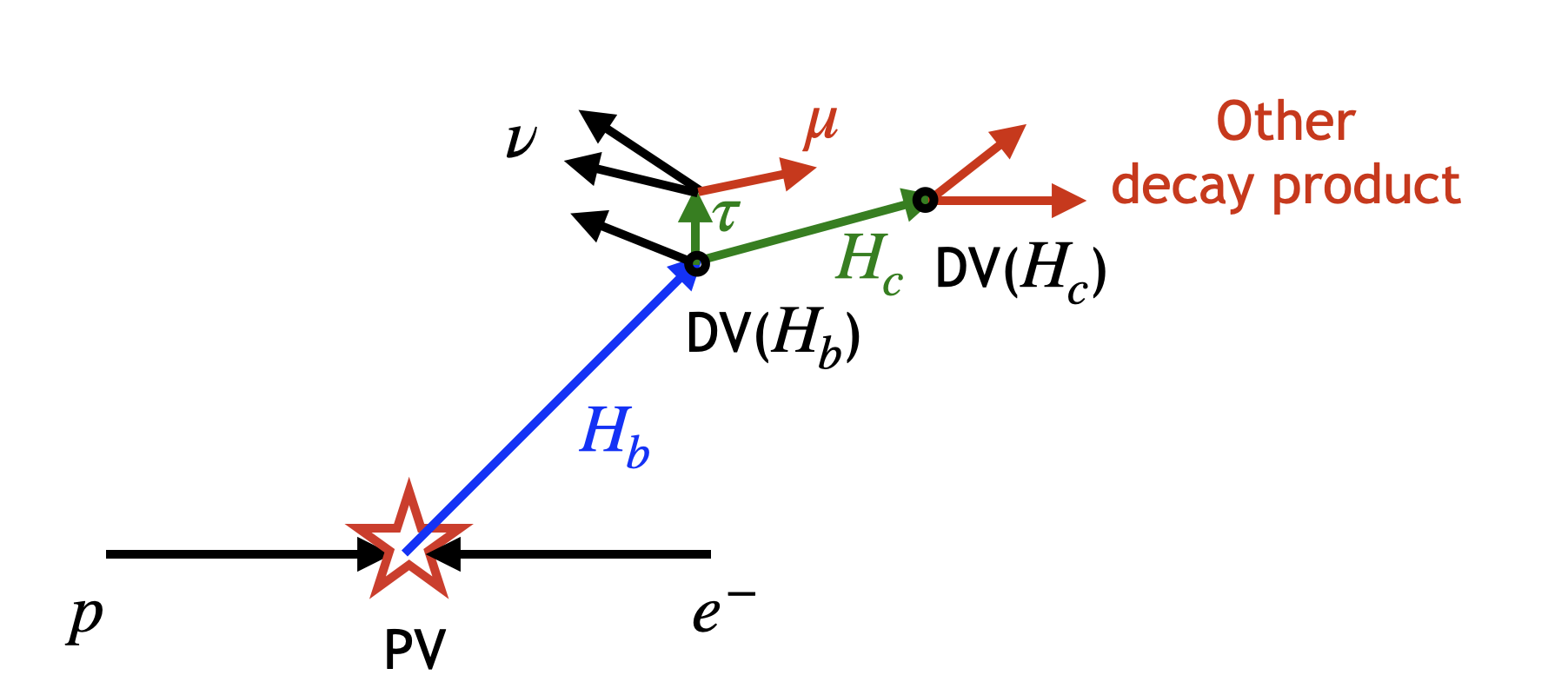}
	\caption{The topology of $H_b\to H_c \tau\nu$ in an electron-proton DIS process at the EIC.} 
	\label{fig:topology}
\end{figure}


We show the topology of $H_b\to H_c \tau\nu$ at the EIC in Fig.~\ref{fig:topology}, and the topology of $H_b\to H_c \mu\nu$ is similar to that of the $\tau$ mode. All signal events are generated using \emph{Pythia8}~\cite{Bierlich:2022pfr}, with the ``Photon-Parton Processes" module utilized to produce $b\bar b$ at the parton level. The showering is carried out with Pythia's default parameters, and we exclusively select the $b$-hadrons of interest, ensuring they decay semi-leptonically into the $c$-hadrons as required. The detector effects are simulated using \emph{Delphes3}~\cite{deFavereau:2013fsa}, following the configuration proposed in Ref.~\cite{Arratia:2021uqr}. For the simplified analysis, we have turned off track vertex smearing effects, as these can be effectively managed during the reconstruction of collision events with a well-designed detector and the implementation of advanced techniques, such as machine learning. For our benchmark simulation, we also assume a perfect muon identification following community discussions of potential detector upgrades~\cite{Zhang:2022zuz}, while neglecting the estimated 5\% pion to muon misidentificaiton rate~\cite{Arratia:2020azl} as its impact on our conclusions would be negligible.

In the background analysis, we focus on the $b$-related hadrons, as discussed in Ref.~\cite{Ho:2022ipo}. Non-$b$ backgrounds are assumed to be distinguishable and separable. All background events are generated inclusively with \emph{Pythia8} and then simulated for detector effects using \emph{Delphes3}. We specifically select backgrounds that share similar decay products with the signal events for further analysis.

\section{Decay processes at the EIC\label{sec: processes}}

\subsection{Measurements of $R_{J/\psi}$}

In this subsection, we examine the measurement of $R_{J/\psi}$ with the decay modes $B_c^+ \to J/\psi(\to \mu^+\mu^-) \mu^+\nu_\mu$ and $B_c^+ \to J/\psi(\to \mu^+\mu^-) \tau^+(\to \mu^+ \nu_\mu\bar\nu_\tau)\nu_\tau$, as proposed by the LHCb collaboration~\cite{LHCb:2017vlu}. For the analysis, both the signal and background events are required to pass the following pre-selection rules:
\begin{itemize}
\item {\bf The muon selection.} We require that the event contains exactly 3 muons, each with $p_T > 0.1$ GeV and the absolute value of their total charge to be unity, $i.e.$, either $\mu^+\mu^-\mu^+$ or $\mu^+\mu^-\mu^-$.

\item {\bf The $J/\psi$ selection.} Among the 3 muons, we require the oppositely charged one forms a decay vertex with at least one of the rest. These 2 muons need to satisfy: each momentum $|\vec p| > 0.5$ GeV, the leading and total transverse momenta both larger than 0.2 GeV, the formed decay vertex at least 0.1 mm away from the primary vertex (PV) and most importantly, their invariant mass falling into the mass window of $J/\psi$, $i.e.$, $|m_{\mu^+\mu^-}-m_{J/\psi}|<27.5$ MeV.

\item {\bf The $B_c^+$ selection.} The space is divided into signal and tag hemispheres with a plane perpendicular to the displacement of the reconstructed $J/\psi$. We require the vertex of $J/\psi$ appearing in the signal hemisphere and so does the third unpaired muon, with its total and transverse momenta greater than 0.5 GeV and 0.2 GeV respectively. Furthermore, we require the 3 muons resulting in an invariant mass below the mass of $B_c^+$.
\end{itemize}

\begin{table}[htbp]
\begin{center}
\fontsize{9pt}{10.8pt}\selectfont 
\begin{tabular}{cccccc}	
	\hline
	Channel
	& Events at the EIC 
	& $N(3\mu)$
	& $N(J/\psi)$ 
	& $N(B_c^+)$ 
	& Total eff.\\
	\hline
	
	$B_c^+\to J/\psi \tau^+\nu_\tau$
	& $1.34\times 10^2$
	& $7.67\times 10^1$
	& $3.74\times 10^1$
	& $3.18\times 10^1$
	& $23.7\%$\\
	
	$B_c^+\to J/\psi \mu^+\nu_\mu$
	& $3.25\times 10^3$
	& $1.97\times 10^3$
	& $9.71\times 10^2$
	& $8.85\times 10^2$
	& $27.2\%$\\
	
	Bkg.
	& $1.11\times 10^7$
	& $5.19\times 10^6$
	& $1.23\times 10^5$
	& $8.85\times 10^4$
	& $0.8\%$\\
 
	\hline
\end{tabular}
\caption{Event yields for preselected signals and backgrounds in the $R_{J/\psi}$ measurement at the EIC. The pre-selection rules are detailed in the text.} \label{tab:Jpsi_yields}
\end{center}
\end{table}

Basically, these pre-selection cuts are proposed following those used in a future Tera-$Z$ factory~\cite{Ho:2022ipo}. Among them, momentum requirements are set at a looser level, since particles at the EIC are less boosted. However, the mass window keeps unchanged, which is exactly identical to that applied in the LHCb collaboration~\cite{LHCb:2017vlu}~\footnote{The strategy has been applied throughout the paper. Particularly, the mass windows used in $R_{D_s^{(\ast)}}$ and $R_{\Lambda_c}$ follow the LHCb collaboration~\cite{LHCb:2020cyw, LHCb:2022piu}.}. The event yields at the EIC are summarized in Table~\ref{tab:Jpsi_yields}, along with a detailed cut-flow table. The signal efficiencies remain above 20$\%$, while the background events are significantly suppressed due to the pre-selection cuts.

\begin{figure}[htbp]
	\centering
 \begin{subfigure}{7cm}
 \centering
	\includegraphics[width=7cm, height=4.33cm]{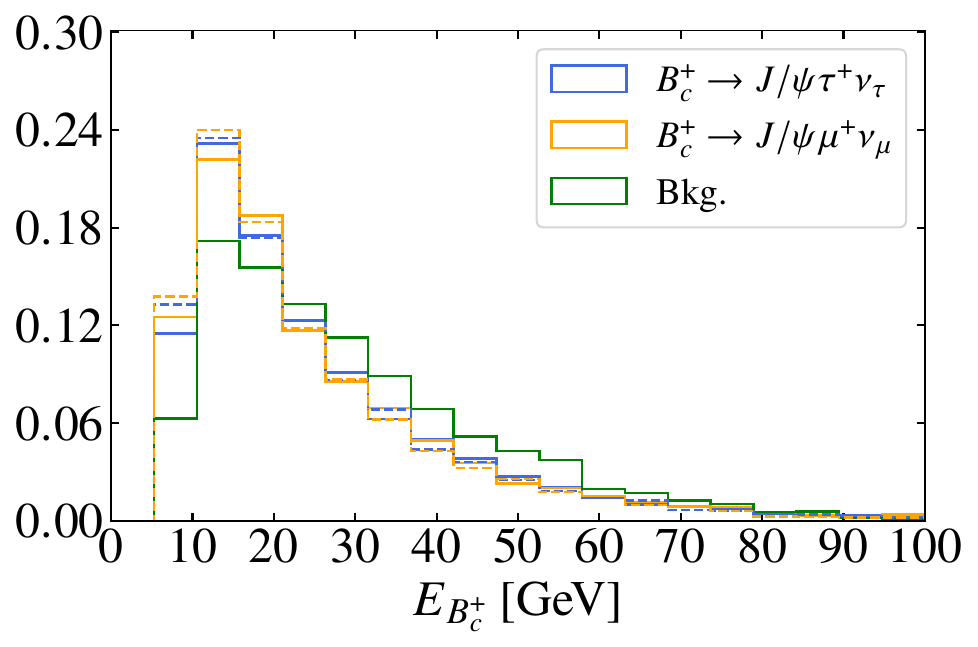}
 \end{subfigure}\hfill
 \begin{subfigure}{7cm}
 \centering
   \includegraphics[width=7cm, height=4.33cm]{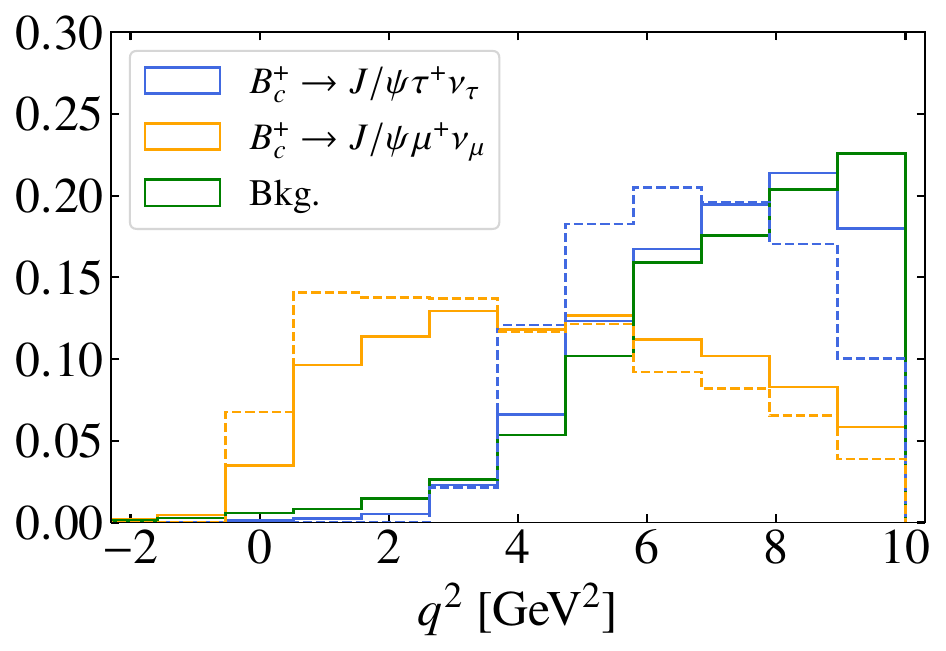}
\end{subfigure}\hfill
\begin{subfigure}{7cm}
\centering
    \includegraphics[width=7cm,height=4.33cm]{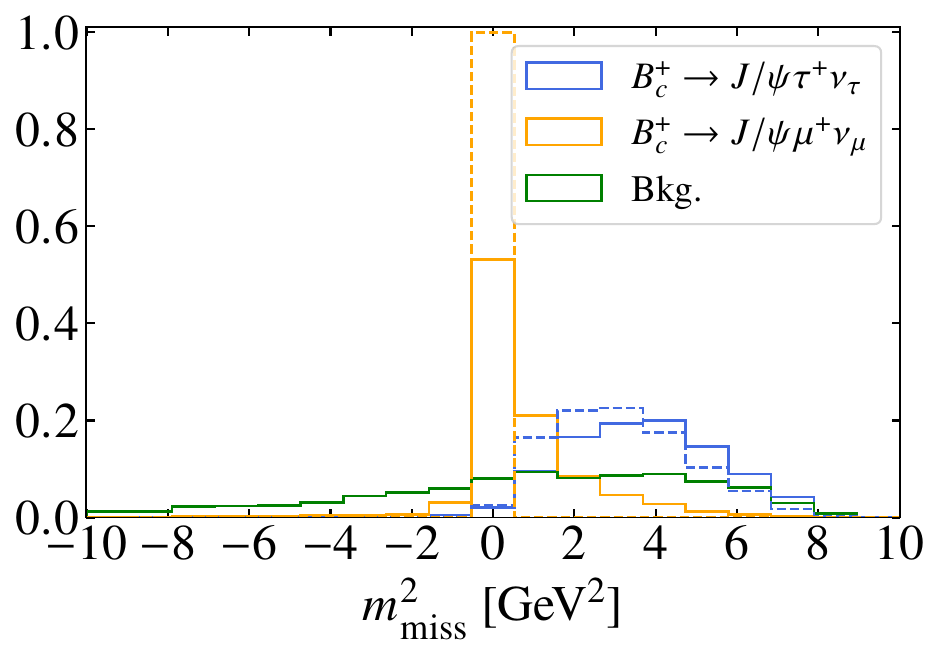}
\end{subfigure}
	\caption{The reconstructed observables of $B_c^+$ are illustrated with solid lines, compared to their truth level predictions, shown with dashed lines. The background distribution at the detector level is also provided for reference.} 
	\label{fig: Bc_Recons}
\end{figure}

Event reconstruction begins with the three detected muons in the final state, followed by sequential reconstruction of the parent particles backward through the decay chain toward the PV. However, the full reconstruction of the $B_c^+$ meson poses significant challenges due to the presence of multiple undetected neutrinos in its decay chain. These neutrinos carry away substantial energy, manifesting as missing transverse energy ($E_\text{T}^\text{miss}$) in the detector. 
To approximate the kinematics of the $B_c^+$ meson, we assume the $J/\psi$ decay vertex, reconstructed from the pair of oppositely charged muons, is treated as approximately coincident with the $B_c^+$ decay vertex due to the short lifetime of the $J/\psi$. Consequently, the momentum of $B_c^+$ is constrained along the direction from the PV to the $J/\psi$ decay vertex.  However, the presence of undetected neutrinos in the decay chain precludes direct application of energy-momentum conservation to fully reconstruct the momentum. Following the method of LHCb collaboration~\cite{LHCb:2015gmp, LHCb:2017vlu}, we estimate the longitudinal momentum component $p_{B_c^+}^{(z)}$ via scaling the visible $3\mu$ system momentum by the mass ratio: 
\begin{equation}\label{eq:visibletoall}
p_{B_c^+}^{(z)}=\frac{m_{B_c^+}}{m_{3\mu}}p_{3\mu}^{(z)}~,
\end{equation}
where $m_{3\mu}$ and $p_{3\mu}^{(z)}$ denote the invariant mass and longitudinal momentum of the reconstructed $3\mu$ system. Figure~\ref{fig: Bc_Recons} compares the reconstructed energy $E_{B_c^+}$ distributions for both signals with their corresponding truth-level distributions. It shows that the robustness of our approach in reconstructing the $B_c^+$ kinematics, with discrepancies well controlled to within approximately $\sim \mathcal O(1)$ GeV.

Similar to the discussion in Ref.~\cite{Ho:2022ipo}, we introduce two Lorentz-invariant observables: $q^2$ and $m^2_{\text{miss}}$, defined as follows:
\begin{align}
q^2 &\equiv (p_{B_c^+} - p_{J/\psi})^2 ~, \label{eq:obs1}\\
m^2_{\text{miss}} &\equiv (p_{B_c^+} - p_{J/\psi} - p_{\mu_3})^2 ~,
\label{eq:obs}
\end{align}
where $\mu_3$ represents the unpaired muon. The distributions of these two observables are shown in Fig.~\ref{fig: Bc_Recons}. Due to the presence of the $\tau$-lepton decays, the decay channel $B_c^+ \to J/\psi(\to \mu^+\mu^-) \tau^+(\to \mu^+ \nu_\mu \bar\nu_\tau)\nu_\tau$ contains additional neutrinos. As a result, this channel is expected to yield larger values for both $q^2$ and $m^2_{\text{miss}}$ compared to other signal. This distinctive feature will play a key role in signal-background separation, as the two signals will contribute as mutual backgrounds in their respective measurements.

To improve the classification of signals and background, we employ the Boosted Decision Tree (BDT) method to assist and optimize the measurement of $R_{J/\psi}$. Specifically, 12 observables, identified as the most important features in ~\cite{Ho:2022ipo}, are selected to train a three-class classifier. These discriminators are listed below, ranked by their average gain across all splits where each feature is used, from most relevant to least relevant:
\begin{itemize}
\item The minimal distance between the $B_c^+$ decay vertex (also the $J/\psi$ decay vertex in our approximation) and the unpaired $\mu_3$ track;
\item The minimal distance between the reconstructed $J/\psi$ trajectory and its closest track;
\item Corrected mass, defined as: $m_{\rm corr} = \sqrt{m^2(J/\psi\mu^+)+p_\perp^2(J/\psi\mu^+)} + p_\perp(J/\psi\mu^+)$ ;
\item The minimal distance between the $\mu_3$ track and its closest track;
\item Distance between the $J/\psi$ decay vertex and the PV;
\item $m_{\rm miss}^2$, as defined in Eq.~\eqref{eq:obs};
\item $J/\psi\mu^+$ momentum transverse to the $B_c^+$ moving direction: $p_{\perp}(J/\psi\mu^+)$;
\item $q^2$, as defined in Eq.~\eqref{eq:obs1};
\item The magnitude of momentum of the reconstructed $J/\psi$: $|\vec{p}_{J/\psi}|$;
\item The magnitude of momentum of the unpaired muon $\mu_3$: $|\vec{p}_{\mu_3}|$;
\item Invariant mass $m_{3\mu}$;
\item The magnitude of momentum of the reconstructed $B_c^+$: $|\vec{p}_{B_c^+}|$.
\end{itemize}

\begin{figure}[htbp]
	\centering
	\includegraphics[width=7cm]{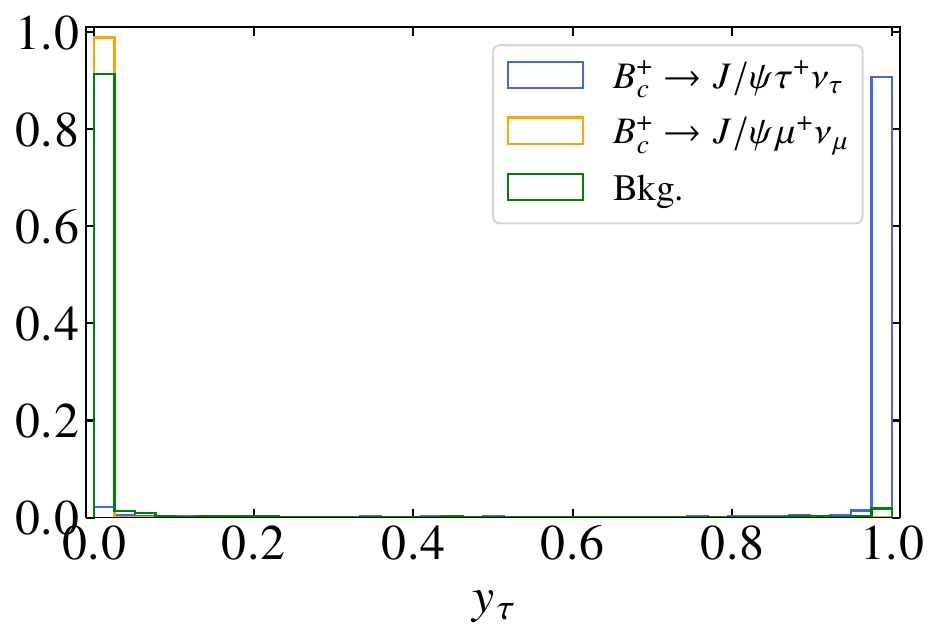}
    \includegraphics[width=7cm]{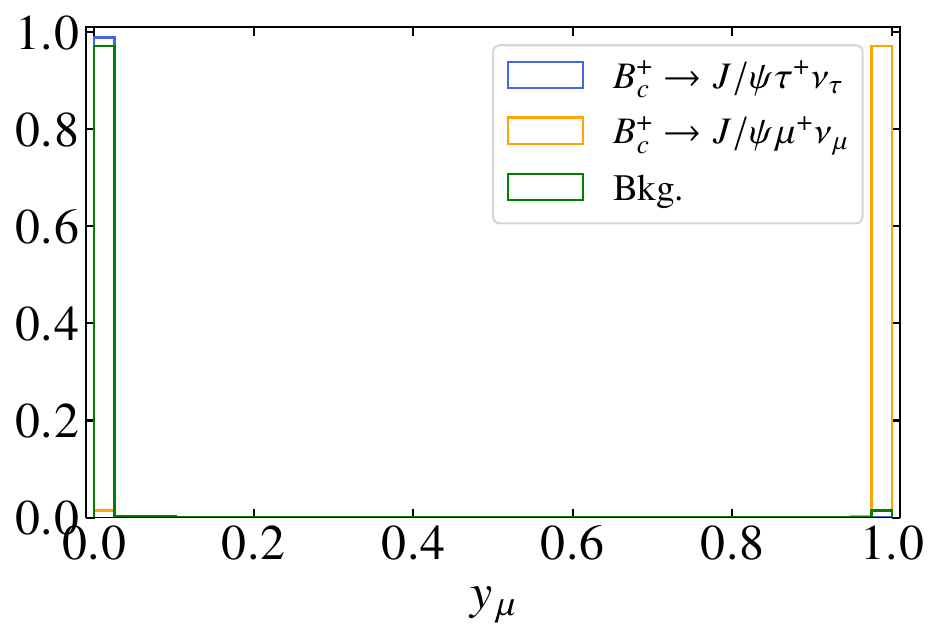}
	\caption{The BDT scores of $B_c^+\to J/\psi \tau^+\nu_\tau$ and $B_c^+\to J/\psi \mu^+\nu_\mu$ are shown as $y_\tau$ and $y_\mu$ respectively. } 
	\label{fig: Bc_BDT}
\end{figure}

The BDT responses for both signals and background are shown in Fig.~\ref{fig: Bc_BDT}. The signal regions  $S_\tau$ and  $S_\mu$ are defined by applying simple cuts to the BDT scores:  
\begin{itemize}
\item $S_\tau$: $y_\tau \geq 0.9$, $y_\mu < 0.2$ and $y_b = 1 - y_\tau -y_\mu < 0.05$;
\item $S_\mu$: $y_\tau < 0.9$, $y_\mu \geq 0.2$ and $y_b < 0.05$.
\end{itemize}
Event yields for these regions are summarized in Table~\ref{tab:BcYield}.  

\begin{table}[htbp]
\begin{center}
\begin{tabular}{cccc}
\hline
	&All & $S_\tau$ & $S_\mu$ \\
	\hline
	$B_c^+ \to J/\psi \tau^+\nu_\tau$ &
    $32$ &
	$29$ &
	$\ll 1$
	\\
	$B_c^+ \to J/\psi \mu^+\nu_\mu$ &	
    $8.85 \times 10^2$ &
	$3$  &
	$8.63\times 10^2$
	\\
	Bkg. & 
    $8.85 \times 10^4$ &
	$1.87\times 10^3$ &
	$1.35\times 10^3$
	\\
	\hline

\end{tabular}
\caption{Event yields in the signal regions $S_\tau$ and $S_\mu$, in terms of the $R_{J/\psi}$ measurement. \label{tab:BcYield}}

\end{center}
\end{table}

The BDT method demonstrates significant success in suppressing backgrounds and substantially enhances the expected statistical significance at the EIC. The expected relative uncertainty (precision) is calculated as $\sqrt{B + S}/S$, where $B$ and $S$ represent the number of background and signal events, respectively. The projected sensitivity for these measurements at the EIC is summarized in Table~\ref{tab:BcRelU}. 
The precision of $R_{J/\psi}$ is primarily limited by the lower reconstruction efficiency for $\tau$-lepton decays compared to muon decays. Under the assumption that statistical uncertainties dominate and systematic contributions remain sufficiently constrained~\footnote{Systematic uncertainties are expected to be significantly reduced in this ratio measurement, as many common sources of uncertainty cancel between the numerator and denominator. Additionally, the large number of background events expected at the EIC enables a detailed understanding of background processes, which in turn allows the relative systematic uncertainty, $\sigma_{\rm sys}$, to be controlled at a low level.}, 
the overall precision of $R_{J/\psi}$ is projected to reach $\mathcal{O}(150)\%$.

\begin{table}[htbp]
\begin{center}
\begin{tabular}{ccccc}	
	\hline
	\multicolumn{2}{c}{$B_c^+\to J/\psi\tau^+\nu_\tau$} &
	\multicolumn{2}{c}{$B_c^+\to J/\psi\mu^+\nu_\mu$} &
	$R_{J/\psi}$\\
	Rel. uncertainty & $S/B$ & 
	Rel. uncertainty & $S/B$ & 
	Rel. uncertainty \\ 
	\hline
	
	 $1.50$ & $1.55\times 10^{-2}$ 
	& $5.45\times 10^{-2}$ & $6.39\times 10^{-1}$  & $1.50$\\
	\hline		
\end{tabular}
\caption{Expected relative uncertainties of measuring $R_{J/\psi}$ at the EIC. 
 \label{tab:BcRelU}}
\end{center}
\end{table}

\subsection{Measurements of $R_{D_s^{(\ast)}}$}

This subsection describes the measurement of $R_{D_s^{(\ast)}}$. For $R_{D_s}$, the signals consist of  $B_s^0\to D_s^- \tau^+\nu_\tau$ and $B_s^0\to D_s^- \mu^+\nu_\mu$, with $D_s^- \to \phi (\to K^+K^-)\pi^-$, while the $R_{D_s^\ast}$ measurement involves a more complex topology: $B_s^0\to D_s^{\ast -} \ell^+\nu_\ell$ followed by $D_s^{\ast -} \to D_s^- \gamma$. Notably, the four signal channels act as mutual backgrounds in their respective measurements and share common final state $K^+K^-\pi^-\mu^+$. To isolate the signals and backgrounds, collision events are pre-selected  with the following cuts:
\begin{itemize}
\item {\bf The $K^+ K^- \pi^- \mu^+$ selection.} We require the selected event share a common decay vertex for the $K^+$, $K^-$ and $\pi^-$, with exactly one moun track having the opposite electric charge to the pion. Additionally, all four tracks should satisfy $p_T > 0.1$ GeV.

\item {\bf The $D_s^-$ selection.} The kaon pair should have an invariant mass satisfying $|m_{K^+K^-} - m_\phi | < 12$ MeV and the decay vertex they form is required to be at least 0.1 mm away from the PV. Furthermore, the $K^+K^-\pi^-$ system is required to have $|m_{K^+K^-\pi^-} - m_{D_s} | < 25$ MeV, where the trajectory of $D_s^-$ is induced according to the reconstructed decay vertex of $K^+K^-\pi^-$ and its momentum $\vec p_{K^+K^-\pi^-}$.

\item {\bf The $B_s^0$ selection.} The space is divided into signal and tag hemispheres with a plane perpendicular to the displacement of the reconstructed $D_s^-$. We require the vertex of $D_s^-$ appearing in the signal hemisphere and so does the muon. The transverse momentum of the muon track should exceed 0.5 GeV. Furthermore, we require the four tracks to form a total invariant mass below the mass of $B_s^0$.
\end{itemize}

The yields at the EIC are summarized in Table~\ref{tab:Ds_yields}, with a detailed cut-flow table. The signals maintain an efficiency of around 5$\%$, while the backgrounds are well suppressed by the pre-selection cuts. 

\begin{table}[htbp]
\begin{center}
\fontsize{9pt}{10.8pt}\selectfont 
\begin{tabular}{cccccc}	
	\hline
	Channel
	& Events at the EIC 
	& $N(K^+ K^- \pi^- \mu^+)$
	& $N(D_s^-)$ 
	& $N(B_s^0)$ 
	& Total eff.\\
	\hline
	
	$B_s^0\to D_s^- \tau^+\nu_\tau$
	& $1.06\times 10^4$
	& $2.14\times 10^3$
	& $8.01\times 10^2$
	& $4.94\times 10^2$
	& $4.66\%$\\
	
	$B_s^0\to D_s^- \mu^+\nu_\mu$
	& $1.55\times 10^5$
	& $3.24\times 10^4$
	& $1.23\times 10^4$
	& $1.01\times 10^4$
	& $6.52\%$\\

 	$B_s^0\to D_s^{\ast -} \tau^+\nu_\tau$
	& $1.78\times 10^4$
	& $3.56\times 10^3$
	& $1.33\times 10^3$
	& $8.12\times 10^2$
	& $4.56\%$\\

 	$B_s^0\to D_s^{\ast -} \mu^+\nu_\mu$
	& $3.45\times 10^5$
	& $7.12\times 10^4$
	& $2.71\times 10^4$
	& $2.22\times 10^4$
	& $6.43\%$\\
	
	Bkg.
	& $9.69\times 10^8$
	& $6.55\times 10^7$
	& $2.32\times 10^5$
	& $1.40\times 10^5$
	& $0.014\%$\\
 
	\hline
\end{tabular}
\caption{The EIC yields for the preselected signals and the backgrounds in the $R_{D_s^{(\ast)}}$ measurement. The pre-selection rules are defined in the text.} \label{tab:Ds_yields}
\end{center}
\end{table}

\begin{figure}[htbp]
	\centering
	\includegraphics[width=7cm,height=4.33cm]{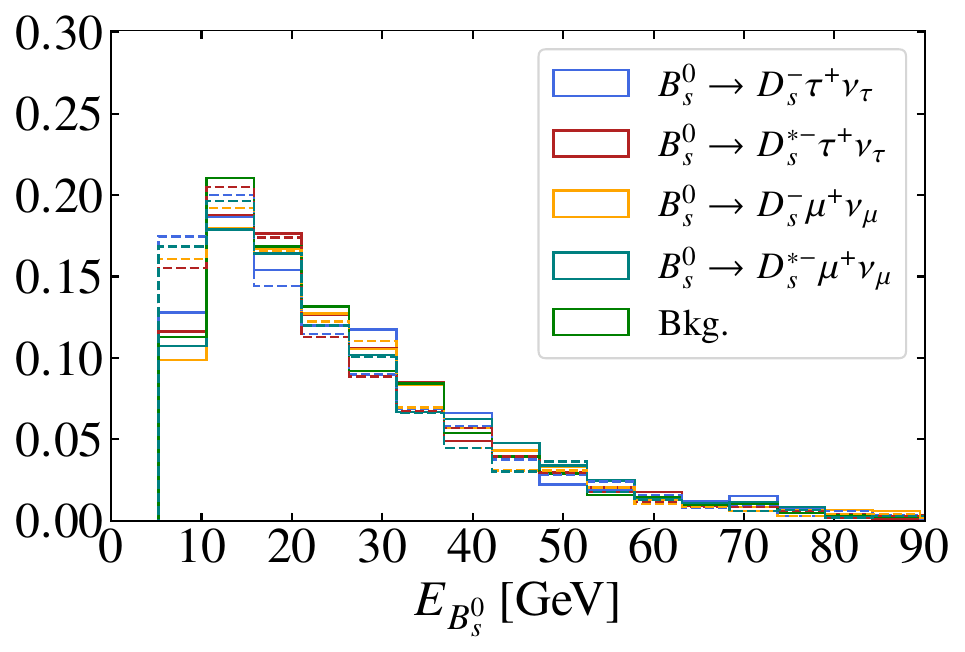}
    \includegraphics[width=7cm,height=4.33cm]{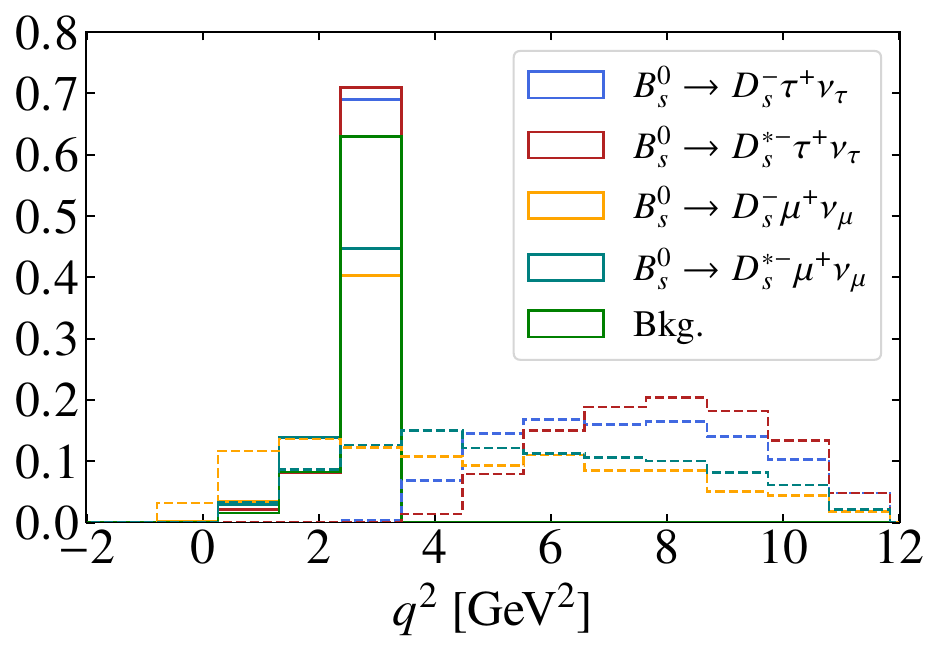}
    \includegraphics[width=7cm,height=4.33cm]{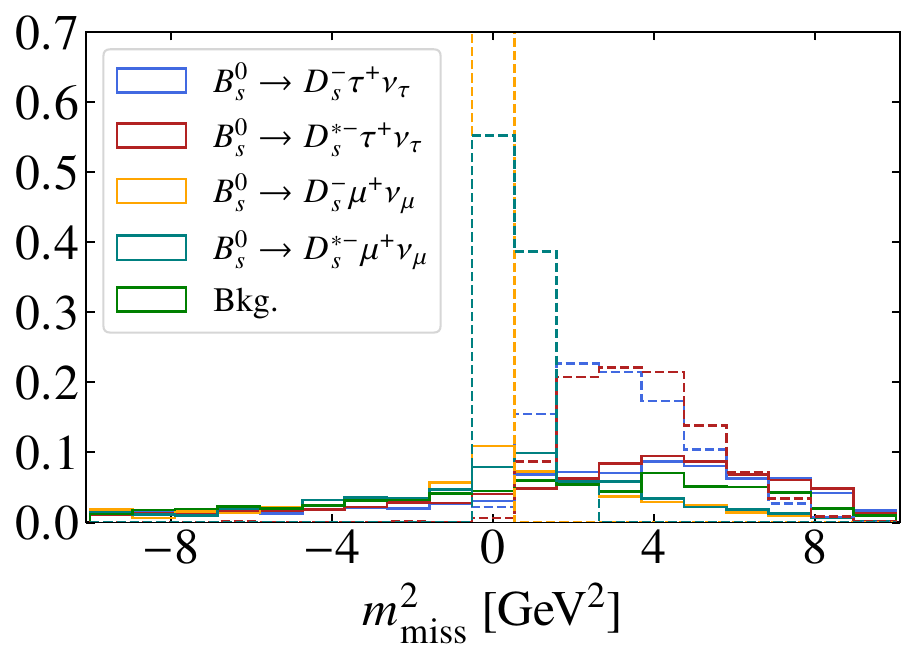}
	\caption{The reconstructed observables of $B_s^0$ are illustrated with solid lines, compared to their truth level predictions, shown with dashed lines. The background distribution at the detector level is also provided for reference.} 
	\label{fig: Bs_Recons}
\end{figure}

The event reconstruction initiates with the $K^+K^-\pi^-$ system. Since the distance traveled by the $\phi$ meson is negligible, it will be ignored in this study. The decay vertex of $D_s^-$ is reconstructed from the 
three decay products, with its trajectory aligned to the momentum $\vec p_{K^+K^-\pi^-}$. The macroscopic distance traveled by the $D_s^-$ is also taken into account. Given the presence of neutrinos, the $B_s^0$ has to be reconstructed in an approximate approach. Specifically, we search along the $D_s^-$ track for the point closest to the muon track and assume it being the $B_s^0$ decay vertex. The momentum of $B_s^0$ aligns with the direction from the PV to the reconstructed $B_s^0$ decay vertex. Its magnitude is estimated from the visible decay products, using the method described in Eq.~\ref{eq:visibletoall}. The reconstruction process is illustrated in terms of the $B_s^0$ energy in Fig.~\ref{fig: Bs_Recons}, along with the corresponding $q^2$ and $m^2_{\rm miss}$, which are also shown in the same figure. The decay topology is complex, and the distance traveled by the $D_s^-$ poses a significant challenge. While the energy of the  $B_s^0$ is well reconstructed, with an error controlled at $\mathcal O(1)$ GeV, the $q^2$ reconstruction suffers from inaccuracies. This differs from the results in Ref.~\cite{Ho:2022ipo}, primarily due to the different reconstruction strategies. At the EIC, the scattering products complicate the reconstruction of the $b\bar b$ system in the center-of-mass frame. As a result, the reconstructed vector direction may misalign with the true orientation, leading to failures in accurately reconstructing $q^2$. Consequently, $q^2$ fails to provide further information to distinguish between different decay processes and has low relevance in the BDT ranking, as demonstrated later.

\begin{figure}[htbp]
\centering
\includegraphics[width=0.7\textwidth]{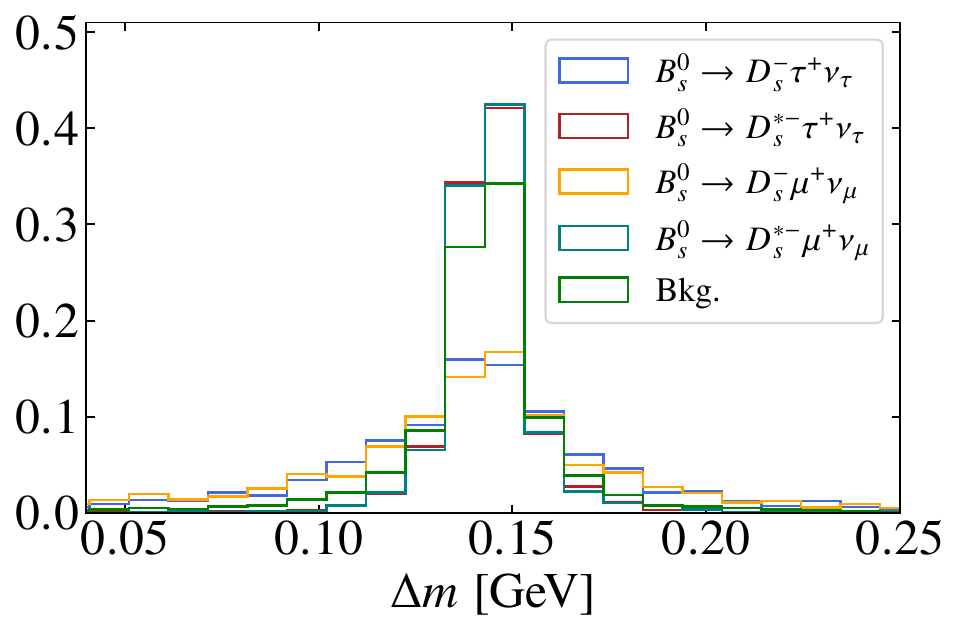}
\caption{The distributions of $\Delta m$ in the measurement of $R_{D_s^{(\ast)}}$ at the EIC.}
\label{fig: dm}
\end{figure}

The distinction between $D_s^-$ and $D_s^{\ast -}$ events is crucial for our study. The difference can be effectively highlighted by the photons recorded in the electromagnetic calorimeter (ECAL). We examine all ECAL photons and identify the one that provides the value of 
$\Delta m \equiv m_{K^+K^-\pi^-\gamma}-m_{K^+K^-\pi^-}$ that is closest to the SM prediction $m_{D_s^{\ast -}} - m_{D_s^-} = 143.8$ MeV. This $\Delta m$ is shown in Fig.~\ref{fig: dm}, serving as one of the most significant discriminators to separate $D_s^-$ and $D_s^{\ast -}$.

The BDT method is employed to optimally separate the signal events from the backgrounds. A five-class BDT is trained using 12 observables. These discriminators are ranked below in order of importance, from most to least significant, based on the average gain across all splits  in which the feature is used:

\begin{itemize}
\item The invariant mass of the $K^+ K^- \pi^- \mu^+$ system: $m_{K^+ K^- \pi^- \mu^+}$;
\item The mass difference: $\Delta m = m_{K^+ K^- \pi^-\gamma} - m_{K^+ K^- \pi^-}$;
\item The minimal distance between the muon track and its closest track;
\item The minimal distance between the $D_s^-$ decay vertex and the muon track;
\item The minimal distance between the deduced $B_s^0$ decay vertex and the muon track;
\item $m_{\rm miss}^2$, similar to Eq.~\ref{eq:obs}, defined as $m_{\rm miss}^2\equiv (p_{B_s^0} - p_{D_s^-} - p_{\mu})^2$;
\item The magnitude of the reconstructed momentum of $D_s^-$: $|\vec p_{D_s^-}|$;
\item $D_s^-\mu^+$ momentum transverse to the $B_s^0$ moving direction: $p_{\perp}(D_s^-\mu^+)$;
\item The distance between the $D_s^-$ decay vertex and the PV;
\item The magnitude of the reconstructed momentum of $B_s^0$: $|\vec p_{B_s^0}|$;
\item Corrected mass, defined as: $m_{\rm corr} = \sqrt{m^2(D_s^-\mu^+)+p_\perp^2(D_s^-\mu^+)} + p_\perp(D_s^-\mu^+)$ 
\item $q^2$, similar to Eq.~\ref{eq:obs1}, defined as $q^2\equiv(p_{B_s^0} - p_{D_s^-})^2$.
\end{itemize}

\begin{figure}[htbp]
	\centering
 
\begin{subfigure}{7cm}
\centering
    \includegraphics[width=7cm,height=4.33cm]{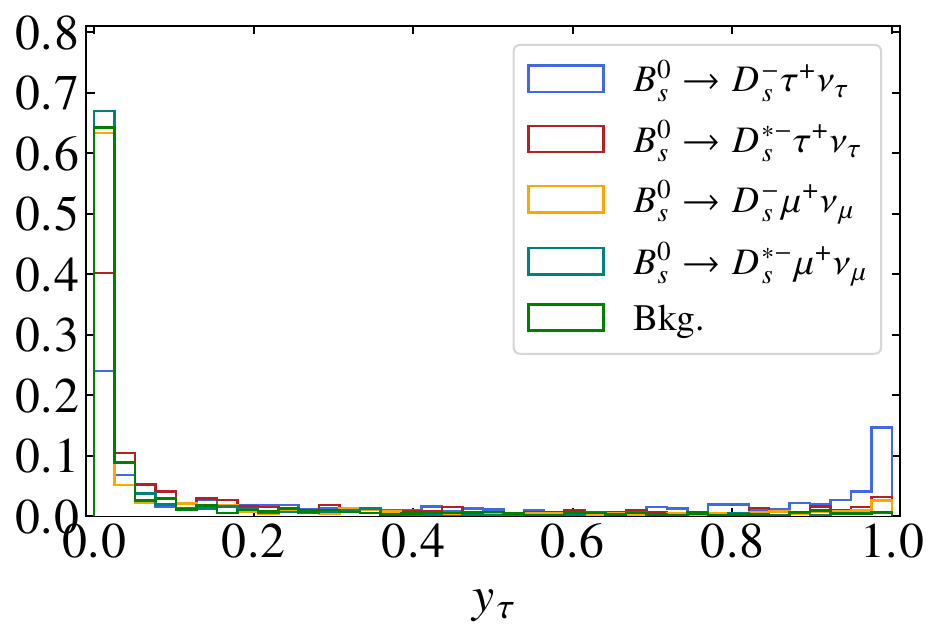}
\end{subfigure}
\begin{subfigure}{7cm}
\centering
    \includegraphics[width=7cm,height=4.33cm]{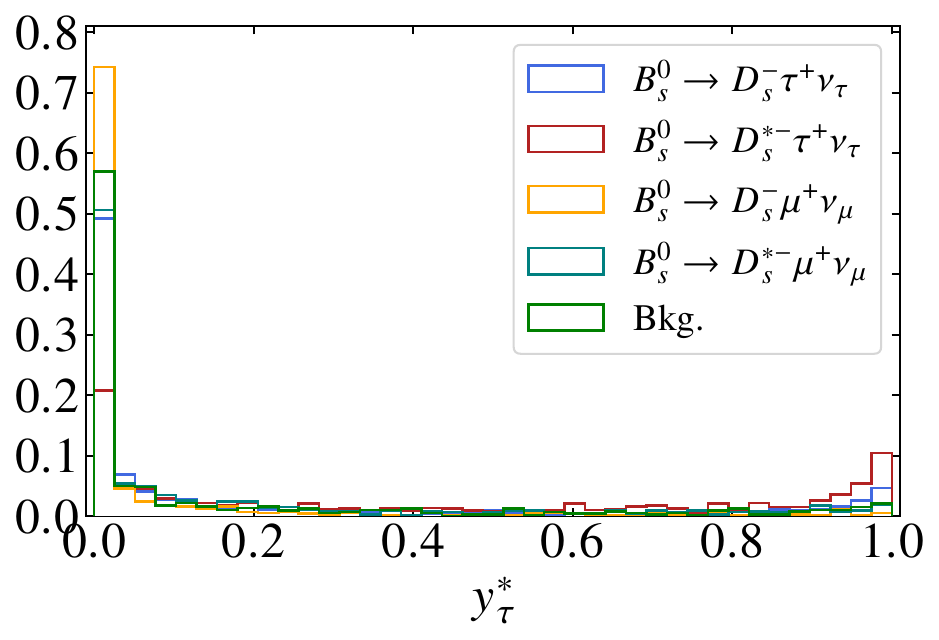}
\end{subfigure}
\begin{subfigure}{0.45\linewidth}
\centering
    \includegraphics[width=7cm,height=4.33cm]{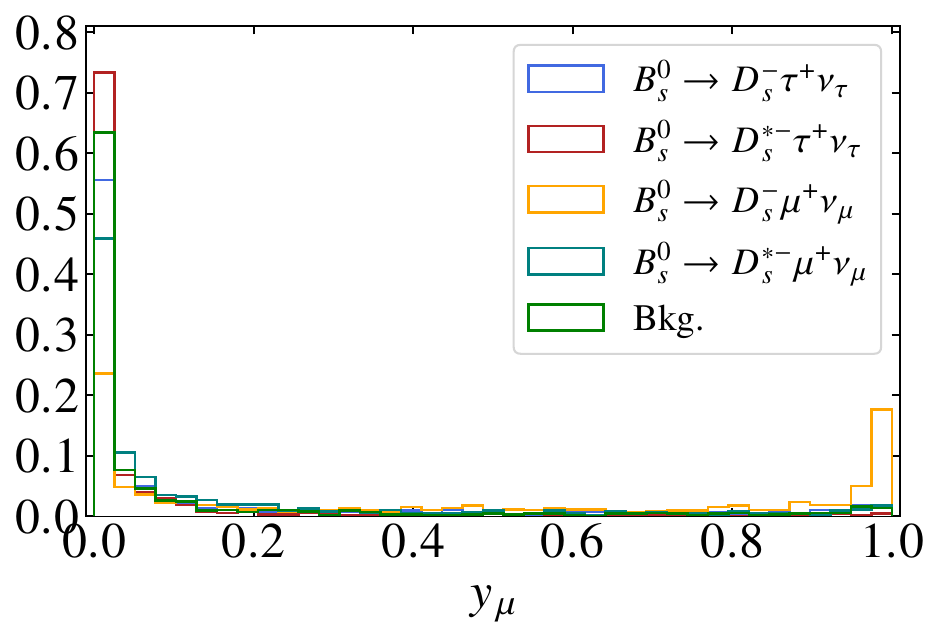}
\end{subfigure}
\begin{subfigure}{0.45\linewidth}
\centering
    \includegraphics[width=7cm,height=4.33cm]{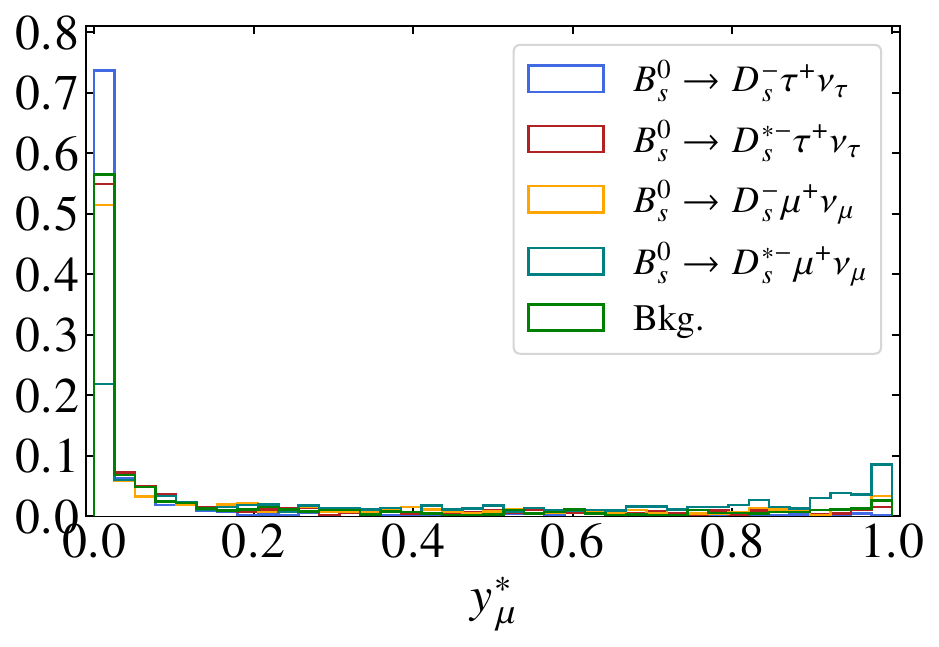}
\end{subfigure}

	\caption{The BDT scores of $B_s^0\to D_s^{(\ast)-} \tau^+\nu_\tau$ and $B_s^0\to D_s^{(\ast)-} \mu^+\nu_\mu$ are shown as $y_\tau^{(\ast)}$ and $y_\mu^{(\ast)}$ respectively. } 
	\label{fig: Bs_BDT}
\end{figure}

The corresponding BDT responses are presented in Fig.~\ref{fig: Bs_BDT}, where $y^{(\ast)}_\tau$ and $y^{(\ast)}_\mu$ corresponding to $B_s^0\to D_s^{(\ast) -} \tau^+\nu_\tau$ and $B_s^0\to D_s^{(\ast) -} \mu^+\nu_\mu$ respectively. The signal regions $S_\tau^{(\ast)}$ and $S_\mu^{(\ast)}$ are defined based on some simple cuts of the BDT scores, as shown below:
\begin{itemize}
\item $S_\tau$: $y_\tau \geq 0.7$, $y_\tau^\ast < 0.3$, $y_\mu < 0.1$, and, $y_\mu^\ast < 0.2$;
\item $S_\tau^\ast$: $y_\tau < 0.7$, $y_\tau^\ast \geq 0.3$, $y_\mu < 0.1$, and, $y_\mu^\ast < 0.2$;
\item $S_\mu$: $y_\tau < 0.7$, $y_\tau^\ast < 0.3$, $y_\mu \geq 0.1$, and, $y_\mu^\ast < 0.2$;
\item $S_\mu^\ast$: $y_\tau < 0.7$, $y_\tau^\ast < 0.3$, $y_\mu < 0.1$, and, $y_\mu^\ast \geq 0.2$.
\end{itemize}
The event yields are summarized in Table~\ref{tab:Bs_yields}, and the expected relative uncertainties that the EIC may achieve are presented in Table~\ref{tab:BsRelU1} and Table~\ref{tab:BsRelU2}. 

\begin{table}[htbp]
\begin{center}
\begin{tabular}{cccccc}
\hline
	&All & $S_\tau$ & $S_\tau^\ast$ & $S_\mu$ & $S_\mu^\ast$\\
	\hline
	$B_s^0 \to D_s^- \tau^+\nu_\tau$ &
    $ 4.94 \times 10^2$ &
	$ 1.56\times 10^2$ &
	$ 1.10 \times 10^2$ &
    $ 1.02 \times 10^2$ &
    $ 3.0 \times 10^1$ 
	\\
 	$B_s^0 \to D_s^{\ast -} \tau^+\nu_\tau$ &
    $ 8.12 \times 10^2$ &
	$ 9.8\times 10^1$ &
	$ 3.73 \times 10^2$ &
    $ 4.8 \times 10^1$ &
    $ 1.16 \times 10^2$ 
    \\
	$B_s^0 \to D_s^- \mu^+\nu_\mu$ &
    $ 1.01 \times 10^4$ &
	$ 7.13\times 10^2$ &
	$ 6.06 \times 10^2$ &
    $ 4.93 \times 10^3$ &
    $ 1.34 \times 10^3$ 
	\\
 	$B_s^0 \to D_s^{\ast -} \mu^+\nu_\mu$ &
    $ 2.22 \times 10^4$ &
	$ 7.84\times 10^2$ &
	$ 3.16 \times 10^3$ &
    $ 3.19 \times 10^3$ &
    $ 7.94 \times 10^3$ 
    \\
	Bkg. & 
    $ 1.40 \times 10^5$ &
	$ 6.59\times 10^3$ &
	$ 2.22\times 10^4$ &
    $ 1.88 \times 10^4$ &
    $ 2.04 \times 10^4$
	\\
	\hline

\end{tabular}
\caption{Event yields in the signal regions $S_\tau^{(\ast)}$ and $S_\mu^{(\ast)}$, in terms of the $R_{D_s^{(\ast)}}$ measurement. \label{tab:Bs_yields}}
\end{center}
\end{table}

\begin{table}[htbp]
\begin{center}
\begin{tabular}{ccccc}	
	\hline
	\multicolumn{2}{c}{$B_s^0\to D_s^- \tau^+\nu_\tau$} &
	\multicolumn{2}{c}{$B_s^0\to D_s^- \mu^+\nu_\mu$} &
	$R_{D_s}$\\
	Rel. uncertainty & $S/B$ & 
	Rel. uncertainty & $S/B$ & 
	Rel. uncertainty \\ 
	\hline
	 $ 5.85\times 10^{-1}$ & $ 1.91\times 10^{-2}$ 
	& $ 3.34\times 10^{-2}$ & $ 2.23\times 10^{-1}$  & $ 5.86\times 10^{-1}$\\
	\hline		
\end{tabular}
\caption{Expected relative uncertainties of measuring $R_{D_s}$ at the EIC. 
 \label{tab:BsRelU1}}
\end{center}
\end{table}

\begin{table}[htbp]
\begin{center}
\begin{tabular}{ccccc}	
	\hline
	\multicolumn{2}{c}{$B_s^0\to D_s^{\ast -} \tau^+\nu_\tau$} &
	\multicolumn{2}{c}{$B_s^0\to D_s^{\ast -} \mu^+\nu_\mu$} &
	$R_{D_s^\ast}$\\
	Rel. uncertainty & $S/B$ & 
	Rel. uncertainty & $S/B$ & 
	Rel. uncertainty \\ 
	\hline
	 $ 4.36\times 10^{-1}$ & $ 1.43\times 10^{-2}$ 
	& $2.18\times 10^{-2}$ & $ 3.63\times 10^{-1}$  & $4.37\times 10^{-1}$\\
	\hline		
\end{tabular}
\caption{Expected relative uncertainties of measuring $R_{D_s^\ast}$ at the EIC. 
 \label{tab:BsRelU2}}
\end{center}
\end{table}

\subsection{Measurements of $R_{\Lambda_c}$}

This subsection focuses on the measurement of $R_{\Lambda_c}$, with the two signal processes being $\Lambda_b^0 \to \Lambda_c^- \tau^+\nu_\tau$ and $\Lambda_b^0 \to \Lambda_c^- \mu^+\nu_\mu$. The $\tau$ lepton is required to decay into a muon and neutrinos, and, the $\Lambda_c^-$ decays through $\Lambda_c^- \to \bar p K^+ \pi^-$, which form a decay vertex. A series of pre-selection cuts have been applied, as summarized below.

\begin{itemize}
\item {\bf The $\bar p K^+ \pi^- \mu^+$ selection.} We require the selected event has $\bar p$, $K^+$ and $\pi^-$ sharing a common decay vertex and there is exactly one muon track with the same electric charge with the kaon. All the four tracks should satisfy $p_T > 0.1$ GeV.

\item {\bf The $\Lambda_c^-$ selection.} The $\bar p K^+ \pi^-$ system forms a common decay vertex, with its distance at least 0.1 mm away from the PV. The invariant mass is required to have $|m_{\bar p K^+ \pi^-}-m_{\Lambda_c}|< 14$ MeV.

\item {\bf The $\Lambda_b^0$ selection.} The space is divided into signal and tag hemispheres with a plane perpendicular to the displacement of the reconstructed $\Lambda_c^-$. We require the vertex of $\Lambda_c^-$ appearing in the signal hemisphere and so does the muon, with its total and transverse momenta exceeding 0.5 GeV and 0.2 GeV, respectively. Furthermore, we require the four tracks to form a total invariant mass below the mass of $\Lambda_b^0$.
\end{itemize}

\begin{table}[htbp]
\begin{center}
\fontsize{9pt}{10.8pt}\selectfont 
\begin{tabular}{cccccc}	
	\hline
	Channel
	& Events at the EIC 
	& $N(\bar p K^+ \pi^- \mu^+)$
	& $N(\Lambda_c^-)$ 
	& $N(\Lambda_b^0)$ 
	& Total eff.\\
	\hline
	
	$\Lambda_b^0\to \Lambda_c^- \tau^+\nu_\tau$
	& $ 1.11\times 10^5$
	& $ 3.13\times 10^4$
	& $ 1.84\times 10^4$
	& $ 1.61\times 10^4$
	& $ 14.5\%$\\
	
	$\Lambda_b^0\to \Lambda_c^- \mu^+ \nu_\mu$
	& $ 1.88\times 10^6$
	& $ 5.70\times 10^5$
	& $ 3.46\times 10^5$
	& $ 2.56\times 10^5$
	& $ 13.6\%$\\
	
	Bkg.
	& $ 2.44\times 10^8$
	& $ 1.64\times 10^7$
	& $ 1.05\times 10^5$
	& $ 9.20\times 10^4$
	& $ 0.038\%$\\
 
	\hline
\end{tabular}
\caption{The EIC yields for the preselected signals and the backgrounds in the $R_{\Lambda_c}$ measurement. The pre-selection rules are defined in the text.} \label{tab:Lambdac_yields}
\end{center}
\end{table}

The yields at the EIC are summarized in Table~\ref{tab:Lambdac_yields}, with a detailed cut-flow table. The signal efficiency is around 14\%, while the backgrounds are significantly suppressed by the pre-selection cuts.

\begin{figure}[htbp]
	\centering
 \begin{subfigure}{7cm}
 \centering
\includegraphics[width=7cm,height=4.33cm]{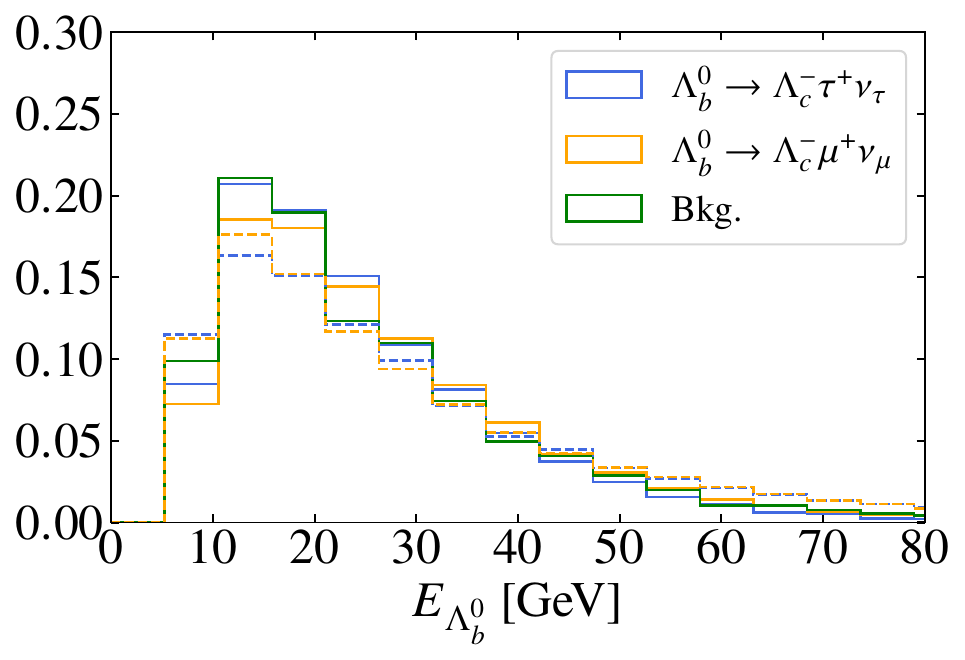}
 \end{subfigure}
\begin{subfigure}{7cm}
 \centering
\includegraphics[width=7cm,height=4.33cm]{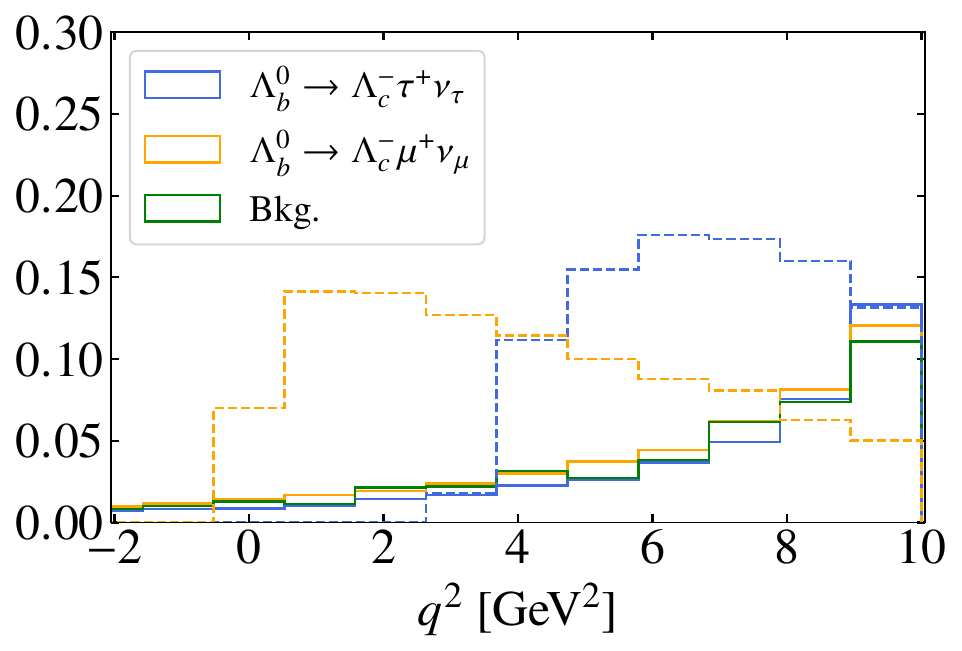}
\end{subfigure}
\begin{subfigure}{7cm}
 \centering
 \includegraphics[width=7cm,height=4.33cm]{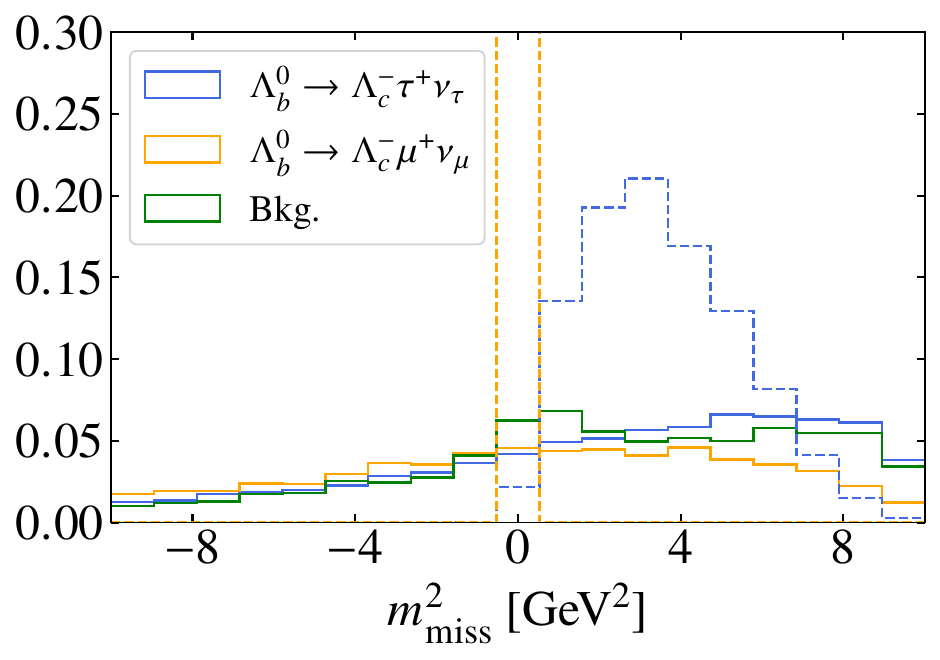}
 \end{subfigure}

	\caption{The reconstructed observables of $\Lambda_b^0$ are illustrated with solid lines, compared to their truth level predictions, shown with dashed lines. The background distribution at the detector level is also provided for reference.} 
	\label{fig: Lambdab_Recons}
\end{figure}

The event reconstruction begins with the $\bar p K^+\pi^-$ system. The decay vertex of $\Lambda_c^-$ is reconstructed according to the three decay products, and its trajectory is taken to be along with $\vec p_{\bar p K^+\pi^-}$. The macroscopic distance traveled by the $\Lambda_c^-$ is also considered. Due to the presence of neutrinos, the reconstruction of the $\Lambda_b^0$ is done approximately.
Concretely, we search along the $\Lambda_c^-$ track for the point closest to the muon track, which is considered as the $\Lambda_b^0$ decay vertex. The momentum of $\Lambda_b^0$ is assumed to align with the direction from the PV to the reconstructed $\Lambda_b^0$ decay vertex, and its magnitude is estimated using the visible decay products, following the method in Eq.~\ref{eq:visibletoall}. 
The reconstruction is illustrated in terms of the $\Lambda_b^0$ energy in Fig.~\ref{fig: Lambdab_Recons}. 
It is evident that the $\tau$-signal reconstruction is relatively worse than that of $\mu$-signal, mainly due to the additional missing components in $\tau$ decays. Nevertheless, the overall reconstruction of $\Lambda_b^0$ energy still shows consistency with the truth level distributions. Additionally, we also introduce $q^2$ and $m^2_{miss}$ as $R_{J/\psi}$ and $R_{D_s^{\star}}$ measurements, which are shown in Fig.~\ref{fig: Lambdab_Recons}. Due to the reconstruction strategy, the reconstruction quality is not optimal, causing these two observables to lose their crucial discriminating power for event classification. In contrast to the magnitude of the decay products, the macroscopic travel distance of the $\Lambda_c^-$ introduces challenges in reconstructing the momentum orientations of the hadron states.

The BDT method is applied to optimally separate the signal events from the backgrounds. A three-class BDT is trained with 12 observables. These discriminators are ranked below, from the most to the least important, according to the average gain across all splits in which the feature is used:
\begin{itemize}
\item The minimal distance between the muon track and its closest track;
\item The minimal distance between the $\Lambda_c^-$ decay vertex and the muon track;
\item $m_{\rm miss}^2$, similar to Eq.~\ref{eq:obs}, defined as $m_{\rm miss}^2\equiv (p_{\Lambda_b^0} - p_{\Lambda_c^-} - p_{\mu})^2$;
\item The invariant mass of the $\bar p K^+\pi^-\mu^+$ system: $m_{\bar p K^+\pi^-\mu^+}$;
\item The magnitude of the muon momentum: $|\vec{p}_{\mu}|$;
\item Corrected mass, defined as: $m_{\rm corr} = \sqrt{m^2(\Lambda_c^-\mu^+)+p_\perp^2(\Lambda_c^-\mu^+)} + p_\perp(\Lambda_c^-\mu^+)$ ;
\item The magnitude of the reconstructed momentum of the $\Lambda_b^0$: $|\vec{p}_{\Lambda_b}|$;
\item $\Lambda_c^-\mu^+$ momentum transverse to the $\Lambda_b^0$ moving direction: $p_{\perp}(\Lambda_c^-\mu^+)$;
\item The minimal distance between the deduced $\Lambda_b^0$ decay vertex and the muon track;
\item $q^2$, similar to Eq.~\ref{eq:obs1}, defined as $q^2\equiv(p_{\Lambda_b^0} - p_{\Lambda_c^-})^2$;
\item The magnitude of the momentum of $\Lambda_c^-$: $|\vec{p}_{\Lambda_c}|$;
\item Distance between the $\Lambda_c^-$ decay vertex and the PV.
\end{itemize}

\begin{figure}[htbp]
	\centering
	\includegraphics[width=7cm]{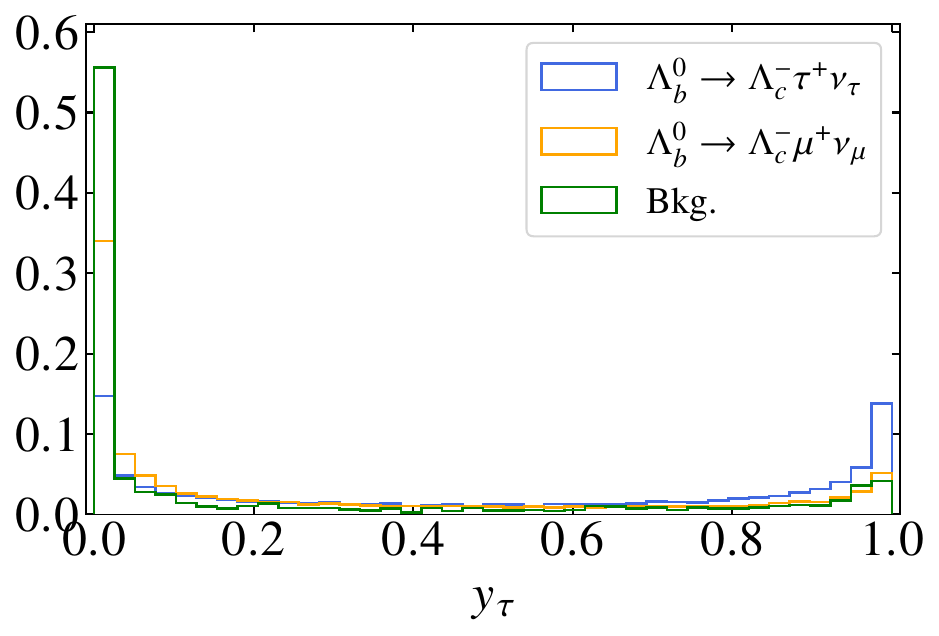}
    \includegraphics[width=7cm]{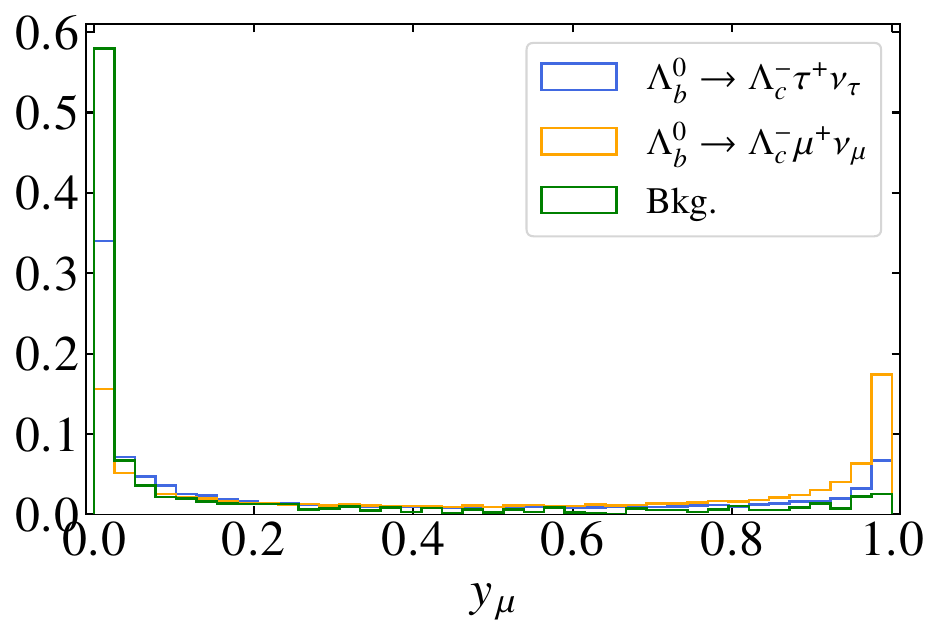}
	\caption{The BDT scores of $\Lambda_b^0\to \Lambda_c^- \tau^+\nu_\tau$ and $\Lambda_b^0\to \Lambda_c^- \mu^+\nu_\mu$ are shown as $y_\tau$ and $y_\mu$ respectively. } 
	\label{fig: Lambdac_BDT}
\end{figure}

The corresponding BDT responses are presented in Fig.~\ref{fig: Lambdac_BDT}. The signal regions $S_\tau$ and $S_\mu$ are defined based on simple cuts of the BDT scores, with $y_\tau \geq 0.2$ and $y_\mu < 0.7$ as $S_\tau$, and, $y_\tau < 0.2$ and $y_\mu \geq 0.7$ as $S_\mu$, respectively. The event yields are summarized in Table~\ref{tab:Lambdab_yields} and the expected relative uncertainties achievable by the EIC are shown in Table~~\ref{tab:LambdabRelU}. Notably, the $\mu$-signal yields are approximately an order of magnitude larger than the backgrounds, making them the dominant contamination when extracting the $\tau$-signal events.

\begin{table}[htbp]
\begin{center}
\begin{tabular}{cccc}
\hline
	&All & $S_\tau$ & $S_\mu$ \\
	\hline
	$\Lambda_b^0 \to \Lambda_c^- \tau^+\nu_\tau$ &
    $ 1.61 \times 10^4$ &
	$ 1.04\times 10^4$ &
	$ 3.31 \times 10^3$
	\\
	$\Lambda_b^0 \to \Lambda_c^- \mu^+\nu_\mu$ &	
    $ 2.56 \times 10^5$ &
	$ 1.00\times 10^5$  &
	$ 1.06\times 10^5$
	\\
	Bkg. & 
    $ 9.20 \times 10^4$ &
	$ 2.75\times 10^4$ &
	$ 1.01\times 10^4$
	\\
	\hline

\end{tabular}
\caption{Event yields in the signal regions $S_\tau$ and $S_\mu$, in terms of the $R_{\Lambda_c}$ measurement. \label{tab:Lambdab_yields}}
\end{center}
\end{table}

\begin{table}[htbp]
\begin{center}
\begin{tabular}{ccccc}	
	\hline
	\multicolumn{2}{c}{$\Lambda_b^0 \to \Lambda_c^-\tau^+\nu_\tau$} &
	\multicolumn{2}{c}{$\Lambda_b^0 \to \Lambda_c^-\mu^+\nu_\mu$} &
	$R_{\Lambda_c}$\\
	Rel. uncertainty & $S/B$ & 
	Rel. uncertainty & $S/B$ & 
	Rel. uncertainty \\ 
	\hline
	 $3.57\times 10^{-2}$ & $8.16\times 10^{-2}$ 
	& $3.26\times 10^{-3}$ & $7.90$  & $3.58\times 10^{-2}$\\
	\hline		
\end{tabular}
\caption{Expected relative uncertainties of measuring $R_{\Lambda_c}$ at the EIC. 
 \label{tab:LambdabRelU}}
\end{center}
\end{table}

\subsection{Consideration of $B_c^+\to \tau^+\nu_\tau$}

In this subsection, we will briefly discuss the annihilation process $B_c^+ \to \tau^+ \nu_\tau$. The potential for studying in this channel is expected to be limited for several reasons. Firstly, the production rate of $B_c^+$ is about two orders of magnitude lower than other $b-$hadrons, which means that the signal yield in this process is comparatively small. Secondly, the reconstruction of the decay is challenging, as there is no straightforward method to trace back and accurately identify the decay vertex of $B_c^+$. One potential solution, demonstrated in future lepton colliders, involves constructing the thrust axis using the visible decay products~\cite{Zheng:2020ult}. However, this method is not applicable at the EIC due to the complexities involved in constructing a valid thrust axis. Consequently, we will not explore the annihilation process at the EIC in this paper. Nonetheless, we believe that advancements in techniques such as machine learning could provide more effective methods for related searches in the future. By comparing the studies of $R_{J/\psi}$, $R_{D_s^{(\ast)}}$, and $R_{\Lambda_c}$ at a future $Z$ factory~\cite{Ho:2022ipo} and at the EIC (as discussed in previous subsections), we conservatively expect the relative uncertainty for the annihilation process to be no worse than 1.5.

\subsection{Comparison with other facilities}

In this subsection, we compare our findings from the EIC with measurements from LHCb and Belle II.  Unlike Belle II, the EIC operates at a collider energy above the $B_c^\pm$ mass threshold, enabling the measurement of $R_{J/\psi}$. However, the EIC cannot  compete with LHCb due to the five orders of magnitude lower production rate. As shown in Ref.~\cite{LHCb:2018roe}, the relative uncertainties of $R_{J/\psi}$ at LHCb are expected to reach $\sim 30\%$ and $\sim 2\%$ for current and future measurements, respectively, which are significantly better than the projected accuracy at the EIC ($\sim 150\%$). The situation for $R_{D_s^{\ast}}$ is different. So far, no results have been reported at colliders for this observable. However, a simple estimation for LHCb Upgrade II suggests that relative uncertainties of about $2.5\%$ could be achieved~\cite{LHCb:2018roe}, which is more than ten times than the EIC. The limitations of the EIC arise from the suppressed $B_s(\bar B_s)$ production and the challenges in $B_s$ reconstruction due to its complex decay topology. To further improve the measurements at the EIC, advancements in $\tau$-tagging technologies and event reconstruction algorithms will be necessary, which are left for future studies. 

The most promising results from the EIC would be in the baryonic sector. Belle II cannot access this channel due to its low energy, while LHCb faces significant QCD backgrounds despite higher $\Lambda_b^0(\bar \Lambda_b^0)$ production rate. As estimated in Ref.~\cite{LHCb:2018roe}, the current and future relative uncertainties at LHCb are $\sim 30\%$ and $\sim 2\%$, respectively. In contract, we demonstrate that the EIC could achieve a relative uncertainty of $\sim 3.58\%$, which is comparable to the upgraded LHCb, thanks to the relatively simpler $\Lambda_b^0$ decay topology and cleaner collision environment.

\section{Effective field theory\label{sec: eft}}

In this section, we adopt the Low-energy Effective Field Theory (LEFT) to perform a model-independent analysis of flavor processes. Within this framework, heavy degrees of freedom are integrated out~\cite{Jenkins:2017jig}, allowing us to systematically incorporate BSM effects through LEFT Wilson coefficients. Notably, we assume that LFU violation occurs only in the third-generation lepton, while the other two generations remain consistent with the SM. We also assume that the measured values for the observables are centered around their SM predictions and the posterior distributions of the Wilson coefficients are then analyzed using Gaussian priors with a Markov-Chain Monte Carlo (MCMC) global fit. Furthermore, we assume that systematic uncertainties can be canceled in the calculation of $R_{H_c}$, and thus we neglect them in this study. The theoretical uncertainties, primarily stemming from form factor parameters, are expected to be well controlled in the future, both theoretically and experimentally.

We consider $b\to c \tau \nu$ transitions within the LEFT framework and limit our analysis to dimension-six (6D) operators. Specifically, the Lagrangian could be parameterized as following:
\begin{equation}
\mathcal{L}^{\rm LE}_{b\to c\tau\nu} =  -\frac{4 G_F V_{cb}}{\sqrt{2}} \bigg [(C_{V_L}^{\tau}|_{\rm SM}+\delta C_{V_L}^{\tau}) O_{V_L}^\tau+ C_{V_R}^{{\tau}} O_{V_R}^\tau+C_{S_L}^\tau O_{S_L}^\tau+C_{S_R}^\tau O_{S_R}^\tau+C_T^\tau O_T^\tau \bigg ]+\rm{h.c.}~,
\label{eq:CCOV}
\end{equation}
where 
\begin{eqnarray}
&&O^\tau_{V_L}=[\bar{c}\gamma^\mu P_L  b ][\bar{\tau}\gamma_\mu P_{L}\nu]~,  \quad O^\tau_{V_R}=[\bar{c}\gamma^\mu P_R b ][\bar{\tau}\gamma_\mu P_{L}\nu]~, 
 \nonumber \\
&&O^\tau_{S_L}=[\bar{c} P_L b ][\bar{\tau} P_{L}\nu]~,  \quad  \quad  \quad
 O^\tau_{S_R}=[\bar{c}P_R b ][\bar{\tau} P_{L}\nu]~,  \nonumber  \\ && O^\tau_{T}=[\bar{c}\sigma^{\mu\nu} b ][\bar{\tau}\sigma_{\mu\nu} P_{L}\nu]~.  
\end{eqnarray}
The superscript ``$\tau$'' indicates the lepton flavor involved, and the subscripts ``$V_L$'', ``$V_R$'', ``$S_L$'', ``$S_R$'', and ``$T$'' correspond to the left- and right-handed vector currents, left- and right-handed scalar currents, and the tensor current, respectively. Notably, $O^\tau_{V_L}$ receives a  non-vanishing SM contribution, with $C_{V_L}^{\tau}|_{\rm SM}=1$, due to the $W$ boson emission. Deviations of these Wilson coefficients from their SM values indicate the presence of BSM effects with LFU violation. The BSM effects in Eq.~\eqref{eq:CCOV} for these observables can be expressed in terms of LEFT Wilson coefficients~\cite{Zheng:2020ult,Ho:2022ipo}.
\begin{equation}
\begin{aligned}
\frac{R_{J/\psi}}{R_{J/\psi}^{\rm SM}} =& ~1.0 + \text{Re} ( 0.12 C_{S_L}^{\tau} + 0.034 |C_{S_L}^{\tau}|^2 - 0.12 C_{S_R}^{\tau} - 0.068 C_{S_L}^{\tau} C_{S_R}^{\tau\ast} + 0.034 |C_{S_R}^{\tau}|^2 \\
& - 5.3 C_{T}^{\tau} + 13 |C_{T}^{\tau}|^2 - 1.9 C_{V_R}^{\tau} - 0.12 C_{S_L}^{\tau} C_{V_R}^{\tau\ast} + 0.12 C_{S_R}^{\tau} C_{V_R}^{\tau\ast} \\
& + 5.8 C_{T}^{\tau} C_{V_R}^{\tau\ast} + 1.0 |C_{V_R}^{\tau}|^2 + 2.0 \delta C_{V_L}^{\tau} + 0.12 C_{S_L}^{\tau} \delta C_{V_L}^{\tau\ast} \\
&- 0.12 C_{S_R}^{\tau} \delta C_{V_L}^{\tau\ast}- 5.3 C_{T}^{\tau} \delta C_{V_L}^{\tau\ast} - 1.9 C_{V_R}^{\tau} \delta C_{V_L}^{\tau\ast} + 1.0 |\delta C_{V_L}^{\tau}|^2  )   \ ,
\end{aligned}
\label{eq:RJpsinumerical}
\end{equation}

\begin{equation}
\begin{aligned}
\frac{R_{D_s}}{R_{D_s}^{\rm SM}}=& ~1.0 + \text{Re}  ( 1.6 C_{S_L}^{\tau} + 1.2 |C_{S_L}^{\tau}|^2 + 1.6 C_{S_R}^{\tau} + 2.4 C_{S_L}^{\tau} C_{S_R}^{\tau\ast} + 1.2 |C_{S_R}^{\tau}|^2 \\
& + 1.4 C_{T}^{\tau} + 1.4 |C_{T}^{\tau}|^2 + 2.0 C_{V_R}^{\tau} + 1.6 C_{S_L}^{\tau} C_{V_R}^{\tau\ast} + 1.6 C_{S_R}^{\tau} C_{V_R}^{\tau\ast} \\
& + 1.4 C_{T}^{\tau} C_{V_R}^{\tau\ast} + 1.0 |C_{V_R}^{\tau}|^2 + 2.0 \delta C_{V_L}^{\tau} + 1.6 C_{S_L}^{\tau} \delta C_{V_L}^{\tau\ast} \\
& + 1.6 C_{S_R}^{\tau} \delta C_{V_L}^{\tau\ast} + 1.4 C_{T}^{\tau} \delta C_{V_L}^{\tau\ast} + 2.0 C_{V_R}^{\tau} \delta C_{V_L}^{\tau\ast} + 1.0 |\delta C_{V_L}^{\tau}|^2  )   \ ,
\end{aligned}
\label{eq:RDsnumerical}
\end{equation}
\begin{equation}
\begin{aligned}
\frac{R_{D_s^*}}{R_{D_s^*}^{\rm SM}}=& ~1.0 + \text{Re}  ( 0.085 C_{S_L}^{ \tau} + 0.026 |C_{S_L}^{ \tau}|^2 - 0.085 C_{S_R}^{ \tau} - 0.052 C_{S_L}^{ \tau} C_{S_R}^{ \tau\ast} \\ 
& + 0.026 |C_{S_R}^{ \tau}|^2 - 4.6 C_{T}^{ \tau} + 15 |C_{T}^{ \tau}|^2 - 1.8 C_{V_R}^{ \tau} - 0.085 C_{S_L}^{ \tau} C_{V_R}^{ \tau\ast} \\
& + 0.085 C_{S_R}^{ \tau} C_{V_R}^{ \tau\ast} + 6.4 C_{T}^{ \tau} C_{V_R}^{ \tau\ast} + 1.0 |C_{V_R}^{ \tau}|^2 + 2.0 \delta C_{V_L}^{ \tau} + 0.085 C_{S_L}^{ \tau} \delta C_{V_L}^{ \tau\ast} \\
& - 0.085 C_{S_R}^{ \tau} \delta C_{V_L}^{ \tau\ast} - 4.6 C_{T}^{ \tau} \delta C_{V_L}^{ \tau\ast} - 1.8 C_{V_R}^{ \tau} \delta C_{V_L}^{ \tau\ast} + 1.0 |\delta C_{V_L}^{ \tau }|^2  )   \ ,
\end{aligned}
\label{eq:RDsstarnumerical}
\end{equation}
\begin{equation}
\begin{aligned}
\frac{R_{\Lambda_c}}{R_{\Lambda_c}^{\rm SM}} =& ~1.0 + \text{Re} ( 0.39 C_{S_L}^{ \tau} + 0.34 |C_{S_L}^{ \tau}|^2 + 0.49 C_{S_R}^{ \tau} + 0.61 C_{S_L}^{ \tau} C_{S_R}^{ \tau\ast} + 0.34 |C_{S_R}^{ \tau }|^2 \\
& + 1.1 C_{T}^{ \tau} + 12 |C_{T}^{ \tau }|^2 - 0.71 C_{V_R}^{ \tau} + 0.49 C_{S_L}^{ \tau} C_{V_R}^{ \tau\ast} + 0.39 C_{S_R}^{ \tau} C_{V_R}^{ \tau\ast} \\
& - 1.7 C_{T}^{ \tau} C_{V_R}^{ \tau\ast} + 1.0 |C_{V_R}^{ \tau}|^2 + 2.0 \delta C_{V_L}^{ \tau} + 0.39 C_{S_L}^{ \tau} \delta C_{V_L}^{ \tau\ast} \\
&+ 0.49 C_{S_R}^{ \tau} \delta C_{V_L}^{ \tau\ast} + 1.1 C_{T}^{ \tau} \delta C_{V_L}^{ \tau\ast} - 0.71 C_{V_R}^{ \tau} \delta C_{V_L}^{ \tau\ast} + 1.0 |\delta C_{V_L}^{ \tau }|^2  )   \ ,
\end{aligned}
\label{eq:RLambdacnumerical}
\end{equation}

\begin{equation}
\begin{aligned}
\frac{\text{BR}(B_c\to \tau\nu)}{\text{BR}(B_c\to \tau\nu)^{\rm SM}}= & ~1.0 + \text{Re} ( 7.1 C_{S_L}^{\tau} + 13 |C_{S_L}^{\tau}|^2 - 7.1 C_{S_R}^{\tau} - 26 C_{S_L}^{\tau} C_{S_R}^{\tau\ast} + 13 |C_{S_R}^{\tau}|^2\\
& - 2.0 C_{V_R}^{\tau} - 7.1 C_{S_L}^{\tau} C_{V_R}^{\tau\ast} + 7.1 C_{S_R}^{\tau} C_{V_R}^{\tau\ast} +1.0 |C_{V_R}^{\tau}|^2 + 2.0 \delta C_{V_L}^{\tau} \\
&+ 7.1 C_{S_L}^{\tau} \delta C_{V_L}^{\tau\ast} - 7.1 C_{S_R}^{\tau} \delta C_{V_L}^{\tau\ast} - 2.0 C_{V_R}^{\tau} \delta C_{V_L}^{\tau\ast} +1.0 |\delta C_{V_L}^{\tau}|^2   )   \ .
\end{aligned}
\label{eq:Bctaununumerical}
\end{equation}

We should note that these observables correspond to different $b\to c$ transitions and have distinct dependencies on the Wilson coefficients, offering the potential to provide constraints from multiple directions in the feature space spanned by these coefficients.  Additionally, LEFT will be matched to the SMEFT~\cite{Grzadkowski:2010es, Azatov:2018knx} to explore its UV completion. It is straightforward to observe that $O^\tau_{V_R}$ in LEFT, does not appear in the matching up to 6D SMEFT operators. Given the irrelevance of $O^\tau_{V_R}$, we therefore set it to zero without any loss of generality.

\begin{figure}[htbp]
\centering
\includegraphics[width=0.9\textwidth]{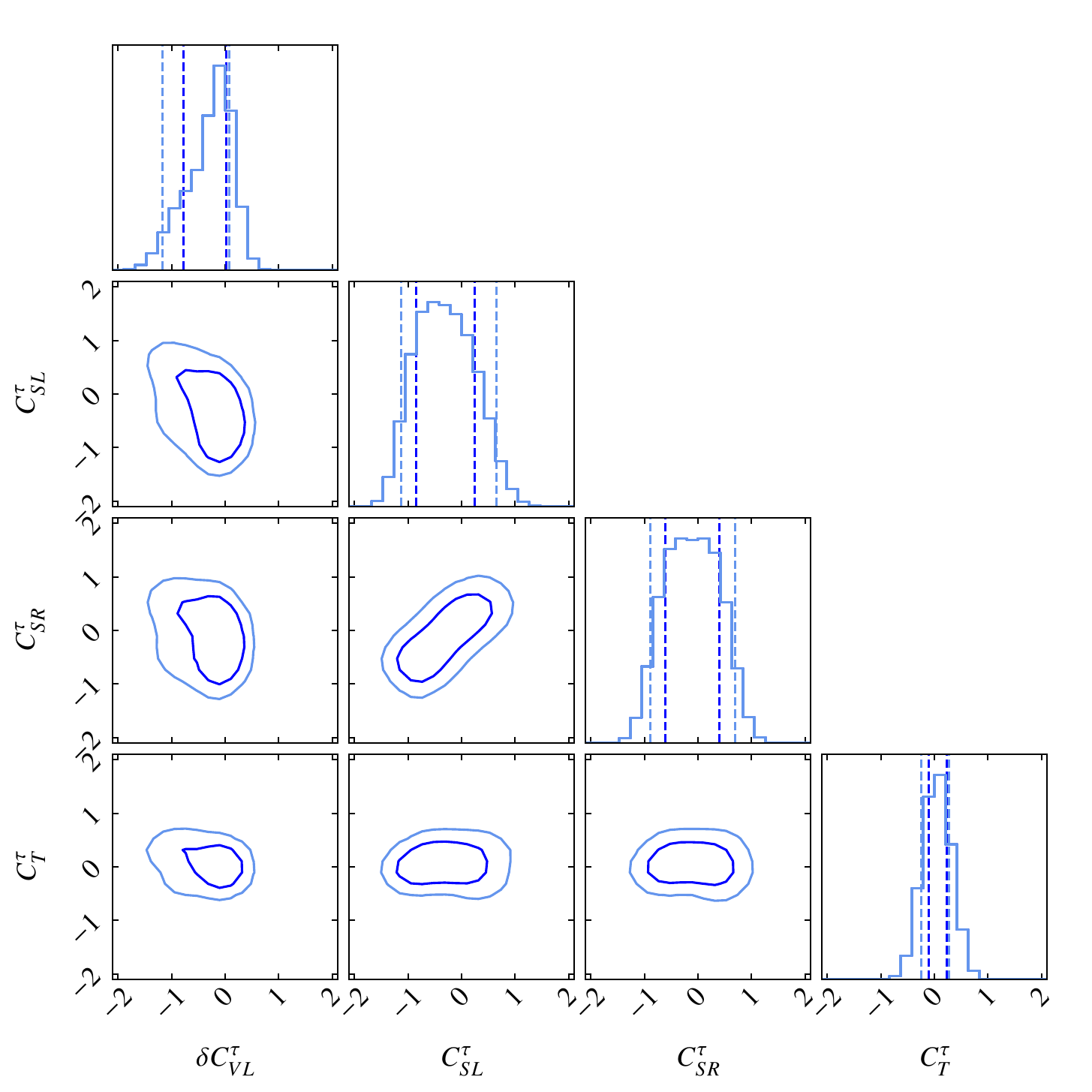}
\caption{2D posterior distributions of the LEFT Wilson coefficients at the EIC, with 68\% (dark blue) and 95\% (light blue) confidence levels.}
\label{fig:LEFT2D}
\end{figure} 

\begin{figure}[htbp]
\centering
\includegraphics[width=0.8\textwidth]{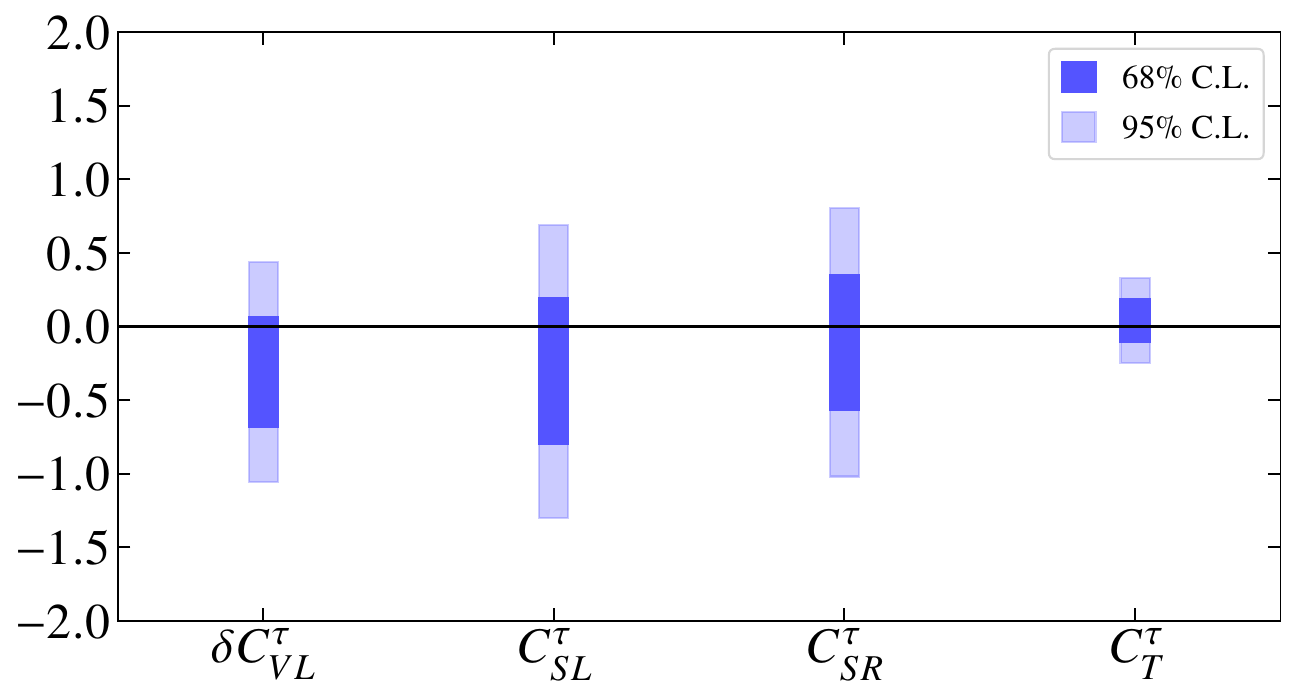}
\caption{1D posterior distributions of the LEFT Wilson coefficients at the EIC, with 68\% (dark blue) and 95\% (light blue) confidence levels.}
\label{fig:LEFT1D}
\end{figure} 

The MCMC fit is performed in a 4D feature space composed of LEFT Wilson coefficients, which are assumed to be real in this work. Gaussian priors are chosen, with the relevant observables centered around their SM values. The posterior distributions are sampled up to $10^5$ points to determine the Wilson coefficients, using the \emph{emcee} package~\cite{Foreman_Mackey_2013}. The 2D posterior distributions are plotted with the \emph{corner} package~\cite{corner} and are shown in Fig.~\ref{fig:LEFT2D}. Additionally, the 1D distributions, obtained by marginalizing over all other degrees of freedom, are presented in Fig.~\ref{fig:LEFT1D}. The correlations between the coefficients are shown as irregular contours in Fig.~\ref{fig:LEFT2D}. These results suggest that all Wilson coefficients are generally constrained to $\sim \mathcal{O}(0.1)$ at the 68\% confidence level at the EIC.

\section{Summary and conclusion\label{sec: conclu}}

Lepton flavor universality (LFU) is one of the cornerstone hypotheses of the SM, and precise tests at colliders are essential to uncover any potential LFU-violating BSM physics. While the EIC has not yet received widespread attention, the substantial $b$-hadron production rates discussed in this paper highlight its potential as a key facility for flavor physics.

We focus on the semileptonic decays of $b$-hadrons, comparing decays involving $\tau$ leptons to those with light flavors. Specifically, we examine $B_c^+ \to J/\psi \ell^+ \nu_\ell$, $B_s^0 \to D_s^{(\ast)-} \ell^+ \nu_\ell$, $\Lambda_b^0 \to \Lambda_c^- \ell^+ \nu_\ell$, and $B_c^+ \to \tau^+ \nu_\tau$. Our findings are compared with the expected reaches at current LHCb, LHCb Upgrade II, and future Tera $Z$-factories, as summarized in Fig. 25 of Ref.~\cite{Ho:2022ipo}.

The performance of $R_{J/\psi}$ and $R_{D_s^{(\ast)}}$ at the EIC is notably worse than projections from LHCb Upgrade II and Tera-$Z$. This discrepancy primarily stems from the limitations in $\tau$ reconstruction efficiency within the current detector configuration. Our analysis is highly dependent on track identification performance, with $\mu$-event reconstruction proving far more successful than $\tau$-event reconstruction. To address uncertainties from $\tau$-decay modes, the implementation of dedicated $\tau$ triggers at the EIC is critical. Advancing detector R\&D and developing more accurate reconstruction strategies, especially tailored to EIC-specific decay topologies, would greatly improve sensitivity by enhancing event classification and suppressing backgrounds.

In contrast, the baryonic $b \to c$ transition offers significant opportunities. For $\Lambda_b^0 \to \Lambda_c^- \ell^+ \nu_\ell$, the EIC’s sensitivity is approximately 30 times worse than that of $Z$-factories but falls between current and upgraded LHCb sensitivities. It's comparable with upgraded LHCb, despite the lower production rate. The EIC benefits from a cleaner environment than LHCb, and its sufficiently high energy makes the study of baryonic modes feasible. With an optimized detector design or reconstruction strategy, the EIC could achieve sensitivity comparable to, or even surpassing, LHCb Upgrade II.

Lastly, we interpret the measured $R_{H_c}$ at the EIC within the framework of LEFT, using an MCMC fit to constrain LEFT Wilson coefficients. Our results suggest that LFU-violating BSM physics could be probed at the EIC when the corresponding Wilson coefficients reach $\sim \mathcal{O}(0.1)$. Importantly, the $b \to c \ell^+ \nu_\ell$ and $b \to s \ell^+ \ell^-$ transitions are strongly correlated due to the $SU(2)_L$ structure in the SM. While this paper primarily focuses on the former channel, a full analysis would also include FCNC processes, such as $b \to s \tau^+ \tau^-$ transitions, which warrant future study due to their relevance in the broader context of flavor physics.

\section*{Acknowledgements}

We thank Jibo He, Tsz Hong Kwok and Lingfeng Li for useful discussions. X.J. was supported in part by the National Natural Science Foundation of China under grant No.~12342502. Y.D. and T.L. were supported by the National Natural Science Foundation of China under grants No. 12175117 and No. 12321005, and Shandong Province Natural Science Foundation under grant No. ZFJH202303. B.Y. was supported in part by the National Science Foundation of China under Grant No.~12422506, the IHEP under Grant No.~E25153U1 and CAS under Grant No.~E429A6M1.

\appendix 
\section{$b$-hadron production at the EIC}
\label{app: bhadrons}

We consider semi-inclusive deep inelastic scattering (SIDIS) processes at the EIC, which are generally represented as:
\begin{equation}
    \label{SIDIS process}
    l(\ell) + N(P) \longrightarrow l\left(\ell^{\prime}\right) + h\left(P_h\right) + X\left(P_X\right),
\end{equation}
where $\ell$, $\ell^{\prime}$, $P$, $P_h$ and $P_X$ denote the four-momenta of the incoming and outgoing leptons, the target nucleon, the detected final-state $b$ hadron and the unobserved remnants, respectively.

After integrating out the momenta of the final-state $b$-hadrons, the differential cross section can be expressed as:
\begin{equation}
    \label{eq: cross section}
    \frac{\mathrm{d}\sigma}{\mathrm{d}x_{\mathrm{B}}\, \mathrm{d}y } = \frac{2\pi\alpha_{EM}^2}{Q^4} s \left[ \left(1 + (1 - y)^2\right)F_2^h\left(x_{\mathrm{B}},  Q^2\right) -y^2 F_L^h\left(x_{\mathrm{B}}, Q^2\right) \right]~,
\end{equation}
with the variables are defined as:
\begin{equation}
    x_{\mathrm{B}} = \frac{Q^{2}}{2P \cdot q}~, \quad Q^{2} = -q^{2}~, \quad y = \frac{P \cdot q}{P \cdot \ell} = \frac{Q^2}{x_{\mathrm{B}} s}~.
\end{equation}
Here, $q = \ell - \ell^{\prime}$ is the momentum of the virtual photon, and $F_{2}^{h}$ and $F_{L}^{h}$ are the structure functions that encapsulate the dynamics of the scattering process. According to the collinear factorization theorem at LO, they can be parameterized as:
\begin{equation}
    \begin{aligned}
        F_{k}^h\left(x_{\mathrm{B}}, Q^2\right) &= \sum  \frac{Q^{2} \alpha_{S} e_{q}^{2} }{4 \pi^{2} m^{2} } 
        \int_{x_{\mathrm{B}}}^{x_{\mathrm{max}}} \frac{\mathrm{d} \tau}{\tau} \, f_{g}\left(\frac{x}{\tau}, Q^{2}\right) c_{k, g}^{(0)}(\eta, \xi)
        \int_{0}^{1} \mathrm{d}z_{h}\, D_{q}^{h}\left(z_h, Q^2\right)~, 
    \end{aligned}
\end{equation}
with $e_q$ as the electric charge of quark flavor $q$, and $m$ is the mass of the quark (or antiquark). Here, $f_{g}(x, Q^{2})$ is the gluon parton distribution function (PDF), which describes the number density of the gluon with momentum fraction $x$ inside the nucleon, and $D_{q}^{h}(z, Q^2)$ is the fragmentation function, describing the number density of a quark fragmenting into the hadron $h$, carrying a fraction $z$ of the quark's momentum. The coefficient functions $c_{k,g}^{(0)}(\eta,\xi)$ ($k = 2, L$) quantify the LO contribution from the photon-gluon fusion process $\gamma g \rightarrow q \bar{q}$. These functions are given by~\cite{Witten:1975bh,Riemersma:1994hv}:
\begin{align}
    c_{L,g}^{(0)}(\eta, \xi) &= \frac{\pi}{2} T_{f} \frac{\xi}{(1+\eta+\xi/4)^3}[ 2(\eta(1+\eta))^{1/2} - \ln\frac{(1+\eta)^{1/2} + \eta^{1/2}}{(1+\eta)^{1/2} - \eta^{1/2}} ]~, \\
    c_{T,g}^{(0)}(\eta, \xi) &= \frac{\pi}{2} T_{f} \frac{\xi}{(1+\eta+\xi/4)^3}[ -2\left\{(1+\eta - \xi/4)^{2} + 1+ \eta \right\}(\frac{\eta}{1+\eta})^{1/2} \\
    &+ \left\{2(1+\eta)^2 + \frac{\xi^2}{8}+ 1+2\eta\right\}\ln\frac{(1+\eta)^{1/2} + \eta^{1/2}}{(1+\eta)^{1/2} - \eta^{1/2}} ]~, \nonumber
\end{align}
where $T_{f} = 1/2$ in $SU(N)$, and $c_{2,g}^{(0)} = c_{L,g}^{(0)} + c_{T,g}^{(0)}$. The scaling variables $\eta$ and $\xi$ are defined as:
\begin{equation}
    \eta = \frac{s}{4 m^{2}} - 1~, \quad \xi = \frac{Q^2}{ m^{2}}~,
\end{equation}
with $s$ is the square of the central energy and $\tau$ satisfies the relation: $\tau = Q^{2}/(Q^{2} + s)$.

To obtain numerical estimations, we use the following inputs:
\begin{itemize}

    \item PDFs: The CT14nlo parameter set~\cite{Dulat:2015mca} for leading-order PDFs,
    
    \item {$B^{(0,+)}$ Fragmentation Function}: The parameterization provided in Ref.~\cite{Kniehl:2007erq}.
    
    \item {$B_c^+$ Fragmentation Function}: The parameterization described in Ref.~\cite{Zheng:2019gnb},
    
    \item {$\Lambda_b^0$ Fragmentation Function}: While an appropriate parameterization is not available, we adopt the approach in Ref.~\cite{Kramer:2018rgb}, assuming that $\Lambda_b^0$ production is similar to $B^{(0,+)}$ mesons, with a normalization adjustment. Specifically, we use the $B^{(0,+)}$ meson fragmentation functions scaled by a factor of $0.5$ for $\Lambda_b^0$.
    
\end{itemize}

The total cross sections for $b$-hadron productions are obtained by integrating Eq.~\ref{eq: cross section} over the relevant phase space at the EIC. The kinematic set-up for the EIC is: 
\begin{equation}
    \begin{aligned}
        &\sqrt{s} = 100\, \mathrm{GeV} ~,\\
        &x_{\mathrm{B}} \in [5\times 10^{-5},\ 0.6]~.
    \end{aligned}
\end{equation}

The estimated $b$-hadron production rates are summarized in Table~\ref{tab:cross_section}. In particular, the cross section for $B_s^0$($\bar{B}_s^0$) is obtained by scaling that of $B^\pm$ meson by a factor of 0.25~\cite{LHCb:2019fns}.
\begin{table}[htbp]
    \centering
    \begin{tabular}{cc}
        \hline
        $b$-hadron &  Total cross section at the EIC [fb] \\ \hline
        $B^{(0,+)} $    & $1.24 \times 10^{6} $ \\
        $B_s^0$   & $3.17 \times 10^{5}$ \\
        $B_{c}^{+} $  & $2.4\times 10^{3}$  \\
        $\Lambda_b^0$   &  $6.2\times 10^{5} $    \\
        \hline
    \end{tabular}
    \caption{The evaluation of $b$-hadron productions at the EIC.}
    \label{tab:cross_section}
\end{table}

\bibliographystyle{JHEP}
\bibliography{EIC}

\end{document}